\documentclass[onecolumn,showpacs,amsmath,
  amssymb,showkeys,floatfix,sort&compress,pre,pdflatex,byrevtex]{revtex4-1}
\usepackage{graphicx}
\usepackage{epstopdf}
\usepackage{bmpsize}
\usepackage{bm}
\usepackage{cancel}
\usepackage{xcolor}
\usepackage{hyperref}
\hypersetup{
  pdftitle={Dynamic Domains of DTS: Simulations of a Spherical Magnetized Couette Flow},
  pdfauthor={E. J. Kaplan},
  colorlinks=true, 
  linkcolor=black,
  urlcolor=black,
  citecolor=black}
\usepackage{natbib}
\usepackage{booktabs}
\usepackage{xspace}

\newcommand{\phiavg}[1]{\left\langle #1 \right\rangle}
\newcommand{\mb}[1]{\mathbf{#1}}
\newcommand{\uv}{\mb{u}}

\newcommand{\bv}{\mb{b}}

\newcommand{\zhat}{\mb{\hat{z}}}
\newcommand{\rhat}{\mb{\hat{r}}}
\newcommand{\phihat}{\mb{\hat{\phi}}}
\newcommand{\Om}{\Omega}
\newcommand{\AOm}{\left|\Omega\right|}
\newcommand{\DOm}{\Delta\Omega}
\newcommand{\Df}{\Delta f}

\newcommand{\curl}{\nabla \times}

\newcommand{\banom}{ \widetilde{\bv}(r, \theta, \phi, t_i)}
\newcommand{\dts}{DTS-$\Omega$\xspace}

\begin{document}

\title{Dynamic Domains of DTS: Simulations of a Spherical Magnetized Couette Flow}
\author{E. J. Kaplan}
\affiliation{Univ. Grenoble Alpes, Univ. Savoie Mont Blanc, CNRS, IRD, IFSTTAR, ISTerre, 38000 Grenoble}
\author{H.-C. Nataf}
\affiliation{Univ. Grenoble Alpes, Univ. Savoie Mont Blanc, CNRS, IRD, IFSTTAR, ISTerre, 38000 Grenoble}
\author{N. Schaeffer}
\affiliation{Univ. Grenoble Alpes, Univ. Savoie Mont Blanc, CNRS, IRD, IFSTTAR, ISTerre, 38000 Grenoble}
\begin{abstract}
  The Derviche Tourneur Sodium experiment, a spherical Couette
  magnetohydrodynamics experiment with liquid sodium as the medium and
  a dipole magnetic field imposed from the inner sphere, recently
  underwent upgrades to its diagnostics to better characterize the
  flow and induced magnetic fields with global rotation. In tandem
  with the upgrades, a set of direct numerical simulations were run to
  give a more complete view of the fluid and magnetic dynamics at
  various rotation rates of the inner and outer spheres. These
  simulations reveal several dynamic regimes, determined by the Rossby
  number. At positive differential rotation there is a regime of
  quasigeostrophic flow, with low levels of fluctuations near the
  outer sphere. Negative differential rotation shows a regime of what
  appear to be saturated hydrodynamic instabilities at low negative
  differential rotation, followed by a regime where filamentary
  structures develop at low latitudes and persist over five to ten
  differential rotation periods as they drift poleward.  We emphasize
  that all these coherent structures emerge from turbulent flows. At
  least some of them seem to be related to linear instabilities of the
  mean flow.  The simulated flows can produce the same measurements as
  those that the physical experiment can take, with signatures akin to
  those found in the experiment. This paper discusses the relation
  between the internal velocity structures of the flow and their
  magnetic signatures at the surface.
\end{abstract}
\maketitle

Spherical Couette flow, a fluid medium sandwiched between two spheres
rotating about a common axis, has a long history in the fluid dynamics
community. Transitions to chaos were investigated in spherical Couette
flow with a stationary outer sphere, with many groups reporting
axisymmetric instabilities, akin to Taylor G\"ortler vortices, that
sometimes broke the equatorial symmetry of the
system \cite{DumasThesis,Egbers.ActaMech.1995,Belyaev.FD.1978,Nakabayashi.PoF.2002,Marcus.JFM.1987}.
Planetlike and starlike systems were studied by rapidly rotating the
outer sphere. At low differential rotation these flows are organized
quasigeostrophically \cite{Proudman.JFM.1956,Stewartson.JFM.1966} with
a large shear across the tangent cylinder, the immaterial cylinder
aligned with the rotation axis and touching the inner sphere's
equator.  The first instability of this Stewartson shear layer is
non-axisymmetric \cite{Hollerbach.JFM.2003,Schaeffer.PoF.2005}.

At moderate counter-rotation, experiments in Maryland reported
inertial modes \cite{Kelley.PRE.2010,Rieutord.PRE.2012} excited by an
over-critical shear across the Stewartson layer.
Wicht \cite{Wicht.JFM.2013} provides a thorough bestiary of the
different regimes in numerical realizations of spherical Couette flow.

Magnetized spherical Couette (MSC) flow, where the fluid is
electrically conducting and a magnetic field is applied from outside
the fluid domain, introduces new dynamics. With an applied dipole
field and a conductive inner sphere, magnetic entrainment yields a
fluid domain rotating faster than the inner sphere. This linear,
theoretical result \cite{Dormy.JFM.2002,Kleeorin.JFM.1997} at low
differential rotation has been extended to nonlinear systems in both
experiments \cite{Brito.PRE.2011} and
simulations \cite{Nataf.PEPI.2008}. Experiments reveal that
fluctuations show up as a succession of broad peaks in the magnetic
frequency spectra \cite{Schmitt.JFM.2008} which correspond to
increasing azimuthal wave numbers. They have been related to linear
modes of the non-linear magnetic base flow, but numerical simulations
suggest that non-linear instabilities could show similar
signatures \cite{Figueroa.JFM.2013}.

\begin{centering}
	\begin{figure}
 	 \includegraphics[width=18cm]{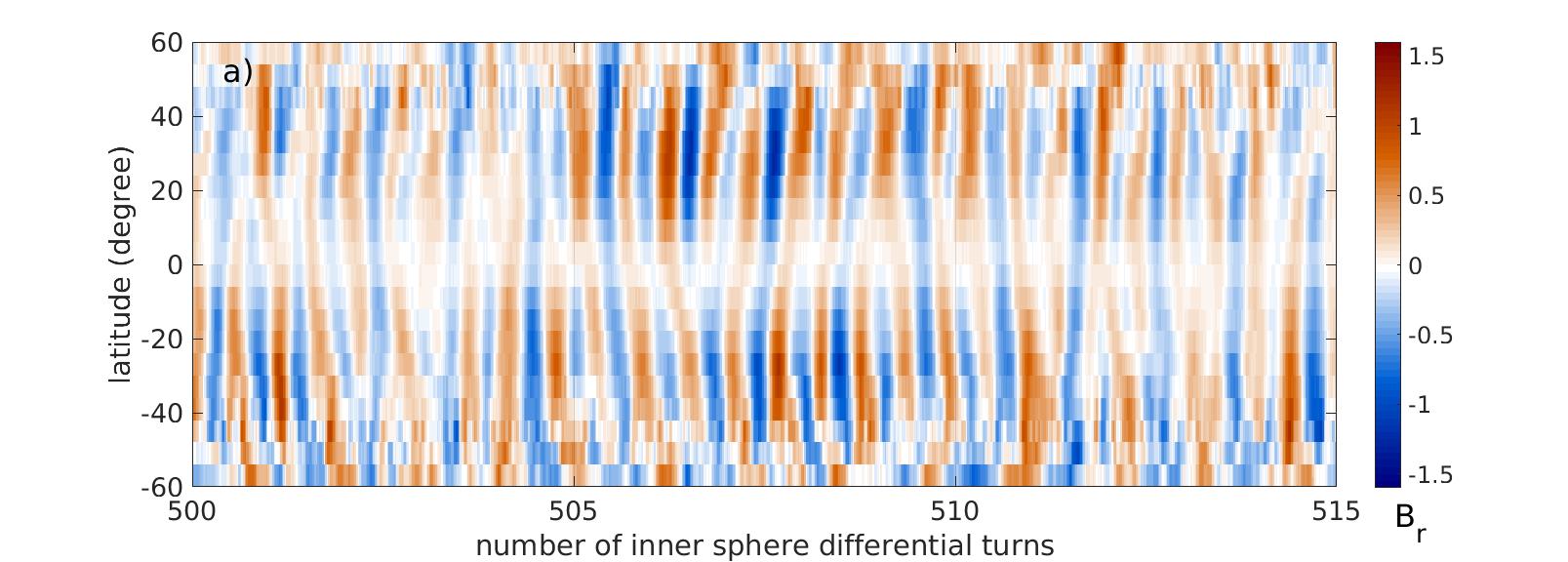}
 	 \includegraphics[width=18cm]{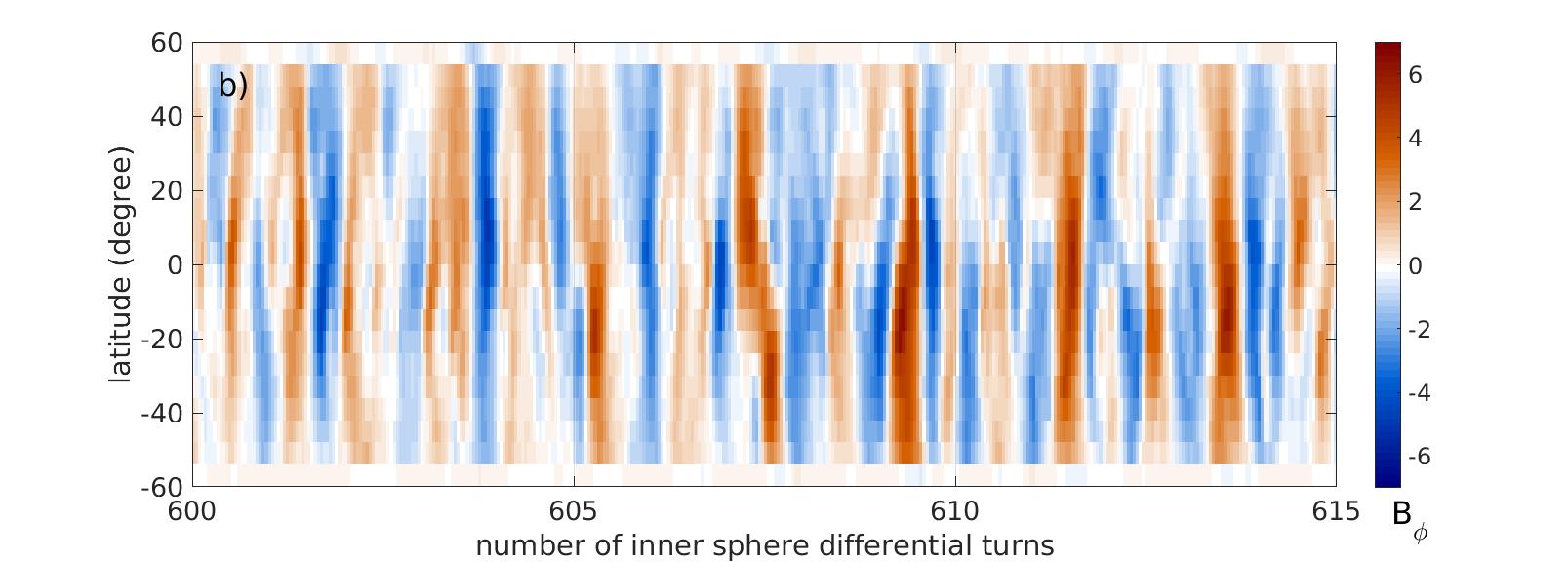}
	  \caption{Time evolution of the induced magnetic field measured along a meridian
	  at the surface of the \dts experiment spinning at 10 rotations per second.
	  The induced magnetic field is expressed as a percentage of $B_o$, the intensity of the imposed magnetic field at the outer sphere's equator.
	  (a) $B_r$ component for $\Delta f = -10$ Hz. (b) $B_\phi$ component for $\Delta f = -20$ Hz.}
 	\label{fig:experiments}
	\end{figure}
\end{centering}

The experimental model for the MSC flow simulated here is the Derviche
Tourneur Sodium experiment (\dts), a $r_i=74$~mm radius inner sphere
containing a strong permanent dipole magnet inside a $r_o=210$~mm
radius outer sphere \cite{Brito.PRE.2011}.  As the name implies, the
fluid medium is liquid sodium.  The shell of the inner sphere is made
of copper (with four times the electrical conductivity of liquid
sodium), while the outer sphere is made of stainless steel (with one
ninth the electrical conductivity of liquid sodium).  The rotation
frequencies of the inner and outer sphere are respectively $f_i$ and
$f_o$, in the lab frame.  A recent set of upgrades to the diagnostics
allows for better diagnosing (both in terms of quality and quantity)
of flows when the outer sphere rotates ($f_o \neq 0$).
As an example, Figure~\ref{fig:experiments} displays the time evolution of
the induced magnetic field at the surface along a given meridian from latitudes
-60 to 60 degrees.
The outer sphere rotation rate $f_o$ is 10 Hz.
Figure~\ref{fig:experiments}a plots the $B_r$ component for $\Delta f = -10$ Hz, while
figure~\ref{fig:experiments}b shows $B_\theta$ for $\Delta f = -20$ Hz, using the spherical coordinate system $(r,\theta,\phi)$, and defining $\Delta f = f_i - f_o$ the differential
rotation rate.

\begin{figure}
  \includegraphics[width=\linewidth]{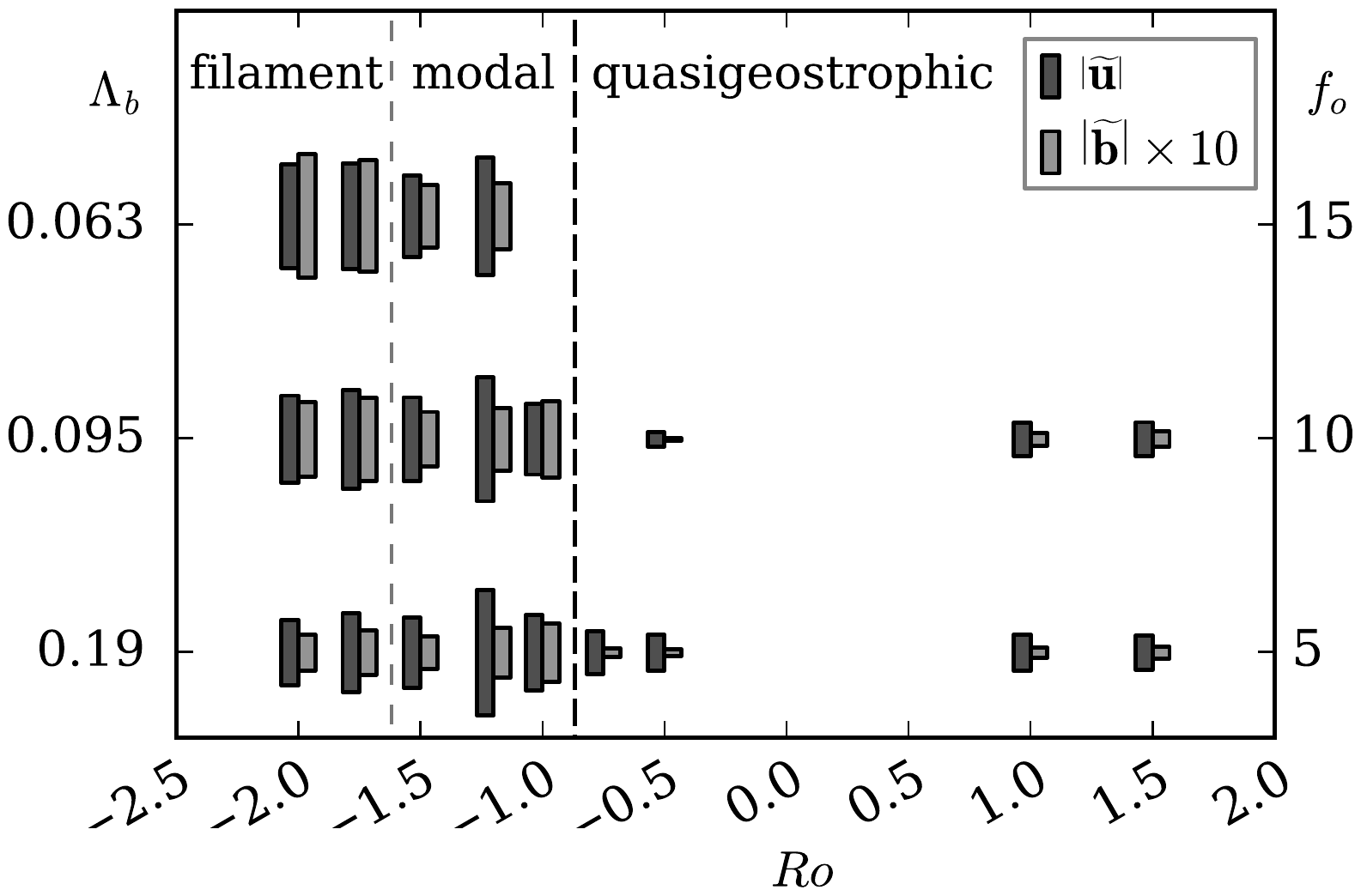} \caption{Dynamic
    regimes of simulations of the DTS device. The heights of the dark
    (light) markers scale linearly with the kinetic (magnetic)
    fluctuation levels. The fluctuation levels are calculated by
    integrating the maps, calculated by
    Eqn.~\ref{eqn:fluctuationprof}, in $r$ and $\theta$. The
    velocities and magnetic fields are both in nondimensional units
    ($1/U_o$ for velocity, $V_A/U_o$ for magnetic fields). The $Em$
    and $\Lambda$ of the simulations are chosen to match those of the
    machine at the various outer rotation rates. (See
    Tab.~\ref{tab:DTSNondim}). The rotation frequency for each
    $\Lambda_b$ value is indicated on the right axis. The $Ek$ of the
    simulations are set to
    10\textsuperscript{-6}. \label{fig:dynregime}}
\end{figure}

In tandem with these upgrades and new run campaigns, a set of direct
numerical simulations (DNS) have been run to predict and interpret the
kinetic and magnetic measurements of the system. The set of DNS
experiments shows three distinct dynamic regimes, though the lines
between them are less than sharp. Figure~\ref{fig:dynregime} shows the
(scaled) energies in the kinetic and magnetic fluctuations for the
full set of DNSs, with indications of where the dynamics, discussed
herein, change from one regime to another.
The Rossby number $Ro = \Delta f/ f_o$ appears as the prime determinant of which regime the
system is in. 
For $Ro \in [-0.5, 1.5]$ the bulk flows appear
quasigeostrophic with weak fluctuations. For $Ro < -1.5$ the flow is
in a turbulent state, characterized by long lived filamentary
structures. For $Ro = -1.2, -1.0$ fluctuations are dominated by large
scale, persistent modes. The determining factor for the kinetic
fluctuation levels changes between regimes. ``Modal'' fluctuation
levels are determined by the $Ro$; ``Filamentary'' fluctuation levels
are determined by the $f_o$, which most likely relates to the
diminished influence of the background magnetic field as the outer
sphere spins faster. The magnetic fluctuation levels are determined by
both $Ro$ and $f_o$, but this only demonstrates that the velocity term
in the magnetic induction equation (Eqn.~\ref{eqn:MagneticInduction}
in Sec.~\ref{sec:numericalmodel}) has increased. The changes in both
fluctuation levels are small over the range of $Ro$s and $f_o$s shown
in ``quasigeostrophic'' regime, so the exact relationship between the
parameters and the fluctuation levels is out of the scope of this
paper.

This paper will proceed as follows. Section~\ref{sec:numericalmodel}
introduces the numerical code used to model the \dts
device. Section~\ref{sec:space} shows the analysis of the computed
velocity and magnetic fields when they are averaged over time and
azimuth. Section~\ref{sec:spattemp} considers the full spatio-temporal
data of the simulations and applies several techniques for analyzing
them. The persistent modes are subject to a principal component
analysis (PCA). The filaments are identified with their magnetic
signatures at the surface. Section~\ref{sec:linear} examines the
linear stability of the mean velocity and magnetic
fields. Section~\ref{sec:time} is a discussion of magnetic
measurements at a single meridian on the outer surface of the
sphere. This is the measurement that is available on the physical \dts
device. Section~\ref{sec:conclusions} are conclusions for this work
and perspectives for the future.
\section{Numerical Model}
\label{sec:numericalmodel}
The dynamical system of the \dts is described by the magnetic
induction and Navier-Stokes equations for an incompressible fluid.
Equations are nondimensionalized by taking as length-scale the radius
of the outer sphere $r_o$ and as characteristic time $\tau \equiv
|\DOm|^{-1}$ -- where $\DOm \equiv 2\pi(f_i - f_o)$ is the angular velocity
difference between the inner and outer sphere. Without loss of
generality, $\DOm  > 0$, while $\Om \equiv 2\pi f_o$ can be either
positive (corotation) or negative (counter-rotation).  After choosing
characteristic velocity $U_o \equiv r_o |\DOm|$, the
nondimensionalized magnetic induction equation reads:
\begin{equation}
  \partial_t \bv = \curl \left(\uv \times \bv\right) +
  \frac{Em}{|Ro|} \Delta \bv \label{eqn:MagneticInduction},
\end{equation}
\noindent where $\uv$ is the incompressible velocity field, $\bv$ is
the magnetic field, $Ro$ is the (signed) Rossby number
$\left(\Omega^{-1}\DOm\right)$, and $Em$ is the magnetic Ekman number
$\left(\eta / r_o^{2} \AOm\right.$, with $\eta$ the magnetic
diffusivity).  The Navier-Stokes equation, in the frame rotating with
the outer sphere, is nondimensionalized as
\begin{align}
  \partial_t \uv + \frac{2}{Ro} \zhat \times \uv &+ \left(\curl 
  \uv\right) \times \uv = \label{eqn:NavierStokes} \\ \nabla p^* &+ \frac{Ek}{|Ro|} \Delta \uv +
  \frac{\Lambda_o Em}{Ro^2} \left(\curl \bv\right) \times \bv \nonumber
\end{align}
\noindent where $Ek$ is the Ekman number $\left(\nu / r_o^{2}
\AOm\right.$ with $\nu$ the viscosity), $\Lambda_o$ is the Elsasser
number at the outer sphere $\left(B_o^2 / \mu_0 \rho \eta
\AOm\right.$, with $B_o$ the intensity of the imposed magnetic field at $r_o$ on the equator, $\mu_0$
the magnetic constant, and $\rho$ the density of the liquid), and
$p^*$ a reduced pressure absorbing all potential forces. The magnetic
field is represented in the code as an Alfv\'en velocity $\left(V_A =
B / \sqrt{\mu_0 \rho}\right)$, and the nondimensional magnetic field
is thus $V_A / U_0$. The values of these parameters in the \dts are
given in Tab.~\ref{tab:DTSNondim}. Because $\Lambda_i$ is always $>>
1$ and $\Lambda_o$ is always $<< 1$, $\Lambda_b$ (the Elsasser number
at $r=(r_i+r_o)/2$) is used to give the best intuition as to the
importance of the magnetic field in a given simulation. The
simulations presented here were designed to match the $Ro$, $Em$, and
$\Lambda_{b,o}$ values of the machine as closely as possible. The $Ek$
is treated as a nuisance parameter and set to 10\textsuperscript{-6};
how and why we treat it that way is handled below.

\begin{table}
  \begin{tabular}{r|ccccc}
    \toprule {} & $Ek$ & $Em$ & $Ro$ & $\Lambda_o$ & $\Lambda_b$ \\ $f_o$ & $\frac{\nu}{r_o^2 \AOm}$ & $\frac{\eta}{r_o^2 \AOm}$
    & $\frac{\Delta\Omega}{\Omega}$ & $\frac{B^2(r=r_o)}{\mu_0 \rho \eta \AOm}$ & $\frac{B^2(r=(r_i+r_o)/2)}{\mu_0 \rho \eta \AOm}$ \\ \hline
    5 Hz & 4.69$\times 10^{-7}$ & 6.38$\times 10^{-2}$ & [-7.0, 5.0] & 1.77$\times 10^{-2}$ & 0.187 \\ 
    10 Hz & 2.35$\times 10^{-7}$ & 3.19$\times 10^{-2}$ & [-4.0, 2.0] & 8.87$\times 10^{-3}$ & 0.095 \\ 
    15 Hz & 1.56$\times 10^{-7}$ & 2.13$\times 10^{-2}$ & [-3.0, 1.0] & 5.92$\times 10^{-3}$ & 0.063 \\ 
    20 Hz & 1.17$\times 10^{-7}$ & 1.60$\times 10^{-2}$ & [-2.5, 0.5] & 4.44$\times 10^{-3}$ & 0.047 \\ 
    \bottomrule
  \end{tabular}
  \caption{Definitions of useful nondimensional parameters for \dts,
    and their values/ranges at various outer sphere rotation
    rates. The two values given for $Ro$ are, respectively, the
    maximum and minimum values attainable for a given $f_o$. The
    values given for $\Lambda_o$ and $\Lambda_b$ are, respectively,
    the Elsasser number at the outer sphere, where the applied field
    is weakest, and at the middle radius sphere, where the dynamics
    actually live. \label{tab:DTSNondim}}
\end{table}

The two dynamical equations are completed by boundary conditions.  The
no-slip boundary condition applies to the velocity field, namely
$\uv(r=r_i) = r \sin\theta \phihat$ and $\uv(r=r_o) = 0$.  The
magnetic field matches with potential fields outside the solid shells
and the conductivity jumps at the solid-liquid boundaries are treated
within the finite difference framework \cite{Cabanes.PRE.2014,
  SHTNS}.  Everything stays spherically symmetric to fit in a
spectral framework.  However, as the goal is to investigate the
properties of a real experiment, the applied magnetic field is chosen
to approximate the field of \dts's inner magnet, rather than a pure
dipole, including magnetic harmonics up to degree 11 and order 6
\cite{Cabanes.PRE.2014}. The inhomogeneous magnetic field is rotated
in phase with the inner sphere.

Because both velocity $\left(\uv\right)$ and magnetic $\left(\bv\right)$ fields are divergence free, our three-dimensional spherical code \texttt{xshells} represents them as solenoidal
vector spherical harmonics
\begin{align}
  \uv(r_i, \theta, \phi) = \sum \limits_{\ell=\left|m\right|}^{\ell_{max}} & \sum \limits_{m=-m_{max}}^{m_{max}} \curl \curl s_{\ell,m}(r_k) Y_\ell^m(\theta, \phi) \rhat \nonumber \\ &+ \curl t_{\ell,m}(r_k) Y_\ell^m(\theta,\phi) \rhat \quad \qquad{\rm and} \nonumber \\
  \bv(r_k, \theta, \phi) = \sum \limits_{\ell=\left|m\right|}^{\ell_{max}} & \sum \limits_{m=-m_{max}}^{m_{max}} \curl \curl S_{\ell,m}(r_k) Y_\ell^m(\theta, \phi) \rhat \nonumber \\ &+ \curl T_{\ell,m}(r_k) Y_\ell^m(\theta,\phi) \rhat,  \label{eqn:specdef}
\end{align}
\noindent where $Y_\ell^m$ are scalar spherical harmonics. The radial
functions are evaluated at every radial grid point, and their
derivatives are taken by second order finite difference.  The forwards
and backwards spherical harmonic transforms are carried out using the
efficient \href{https://bitbucket.org/nschaeff/shtns}{\texttt{SHTns}
  library} \cite{SHTNS}. \texttt{xshells} performs the time stepping
of Eqn.~\ref{eqn:NavierStokes} in the fluid spherical shell, and the
time stepping of Eqn.~\ref{eqn:MagneticInduction} in both the
conducting walls and the fluid. It uses a semi-implicit
Crank-Nicholson scheme for the diffusive terms, while all other terms
are handled by an Adams-Bashforth scheme (second-order in time).  The
time, space, and spectral resolutions of the simulations vary most
strongly with $Ek$ for numerical stability.  The resolutions for the
runs presented here are specified in appendix
\ref{app:SimulationResolution}.

\begin{centering}
  \begin{figure}
    \includegraphics[width=\linewidth]{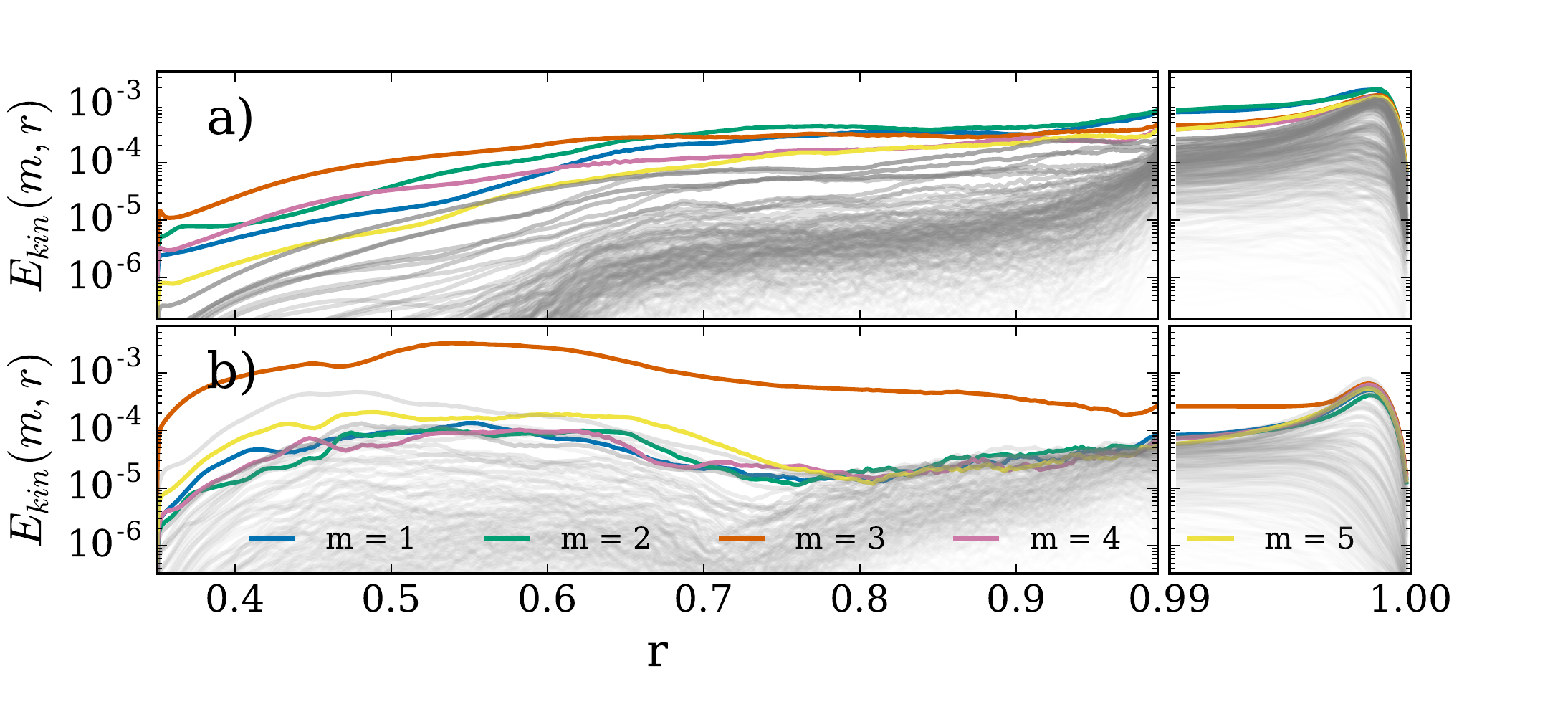}
    \caption{Colormaps showing kinetic energy in each $m>0$ azimuthal
      wavenumber as a function of spherical radius for (a) $Ro = -2.0,
      \Lambda_b = 0.095$ and (b) $Ro = -1.0, \Lambda_b = 0.095$. The
      $m \in [1, 5]$ are colored, the others have their opacity set as
      a function of the largest value in the bulk flow. A zoom of the
      last 0.01 is shown on the right. \label{fig:spectra}}
  \end{figure}
\end{centering}

Preliminary simulations in the range $10^{-5} > Ek \geq 10^{-7}$
suggested that $Ek$ serves to set the spatial resolution required for
numerical stability, but does not drastically affect the dynamics of
the large eddies in the bulk flow, as long as $Ek$ is small enough.
The induced magnetic spectra were similarly agnostic to $Ek$.  Thus we
make two assumptions in the interest of reducing the necessary
spectral resolution.  The first is that $Ek = 10^{-6}$ adequately
represents the large scale component of the flow.  The second is that
a hyperviscosity \cite{Glatzmaier.PEPI.1995} adequately models the
turbulent cascade to small scales in the Ekman layer.  To justify
this, Fig.~\ref{fig:spectra} shows the energy in each azimuthal
harmonic $m$ as a function of radius. The higher order harmonics are
only really visible in the outermost 0.005 of the sphere. The
underlying assumption of hyperviscosity is that, for a turbulent
cascade, there is a net transfer of energy to smaller and smaller
scales through a Kolmogorov cascade, and that this energy transfer can
be modeled as an effective viscosity that increases in value at
smaller and smaller scales. Our simulations use a modified version of
the hyperviscosity \cite{Nataf2015}, that only applies to the angular
part of the Laplace operator and to the last 20\ 
$\left(\ell_0 > 0.8 \ell_{max}\right)$ according to
\begin{equation}
  \nu_{\ell > \ell_0} = \nu_0 \times h^{\left(\ell-\ell_0\right) /
    \left(\ell_{max}-\ell_0\right)}
\end{equation}
\noindent where $\nu_0$ is the base viscosity and $h$ is the ratio
between the viscosity at $\ell_{max}$ and $\nu_0$, here chosen to
be 1000.

In the majority of cases at $\Lambda_b = 0.19, 0.095$, the simulations were
initialized from a previous set of simulations with identical $Ro$ and
slightly different $\Lambda$ and ${\rm E_m}$. The simulations at $\Lambda_b =
0.063$ were initialized from runs at the same $Ro$ at $\Lambda_b =
0.095$. The runs at $Ro = -0.50, -0.75$, and $-1.75$ were seeded off
simulations at $Ro = -1.00, -0.50$, and $-1.50$ respectively. In all
cases, the analyses are carried out after waiting 40 differential
rotation periods for the new simulation to settle. Unless otherwise
mentioned, statistics and analyses are reckoned over 60 differential
rotation periods.

  
\section{Time averaged spatial data}
\label{sec:space}

The diagram of Fig.~\ref{fig:dynregime} shows the kinetic and magnetic
fluctuation levels of the simulations. These fluctuations are defined
in a frame of reference rotating with the inner sphere
\begin{align}
  \overline{\bv}(r, \theta, \phi) &= \frac{1}{N} \sum
  \limits_{i=1}^{N} \bv(r, \theta, \phi, t_i){\rm, \;and}\\
 \banom &= \bv(r, \theta, \phi, t_i) - \overline{\bv}(r, \theta,
 \phi).
\end{align}
\noindent The corotation is necessary for the magnetic fluctuations
because the applied field is three dimensional. A two dimensional
map of the kinetic or magnetic energy densities can be found by integrating
\begin{equation}
  \phiavg{\widetilde{\bv}\cdot\widetilde{\bv}}(r, \theta) =
  \frac{1}{N} \sum \limits_{i=1}^N \frac{1}{2\pi} \int
  \limits_0^{2\pi}d\phi \widetilde{\bv}\cdot \widetilde{\bv}(r,
  \theta, \phi, t_i). \label{eqn:fluctuationprof}
\end{equation}
\noindent these maps, when integrated in $r$ and $\theta$, give the
values represented in Fig.~\ref{fig:dynregime}. 

\begin{figure}
  \includegraphics[width=0.8\linewidth]{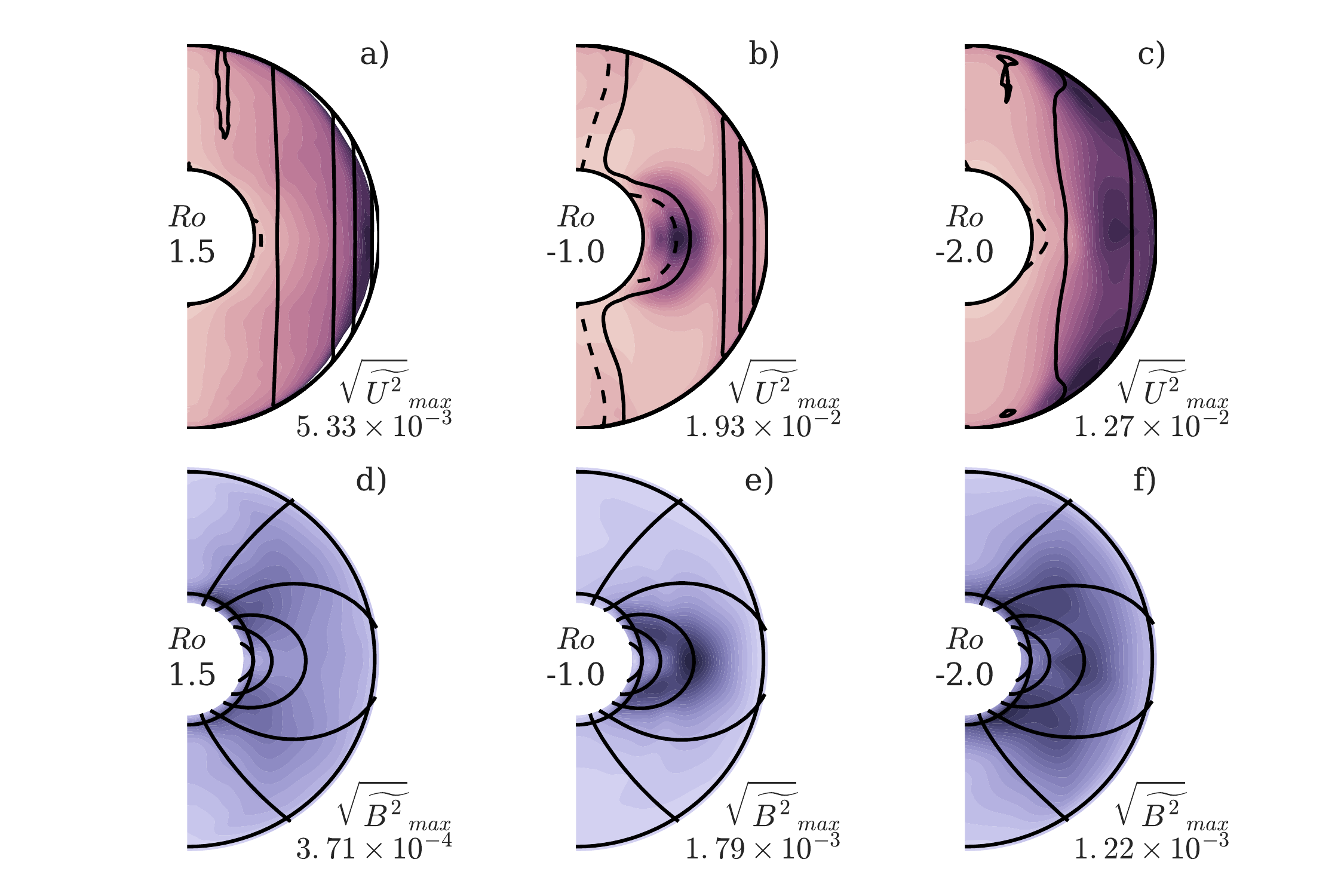}
  \caption{Colormaps showing fluctuation levels of the magnetic and velocity fields. The
    first row, (a, b, and c), shows the kinetic fluctuations, the
    second row, (d, e, and f), shows the magnetic fluctuations. The
    numbers on the lower right indicate the maximum value of the
    energy density (excluding the Ekman layer). The contour lines in
    the first row represent isocontours of angular momentum of the
    bulk fluid. Inside the dashed line the bulk fluid is rotating
    faster than the inner sphere in the reference frame of the outer
    sphere.  The contour lines in the second row are field lines of
    the mean poloidal magnetic field. Each column presents a set of
    parameters firmly in the quasigeostrophic (a and d), modal (b and
    e), and filamentary (c and f) regimes. All of the fluctuation
    levels are calculated for $\Lambda_b = 0.095$. \label{fig:spaceavg}}
\end{figure}

The maps of kinetic and magnetic fluctuation levels for several flows
are shown in Fig.~\ref{fig:spaceavg}. The top row shows the kinetic
fluctuation levels with isocontours of the angular velocity
overplotted; the dashed lines indicate the region of
superrotation ({\it i.e.}, where the fluid rotates with an angular velocity larger than that of the inner sphere \cite{Brito.PRE.2011}). The bottom row shows the magnetic fluctuation levels with fieldlines of the poloidal magnetic field
overplotted. Maps for $Ro = 1.5, -1.0,$ and $-2.0$, all at $\Lambda_b =
0.095$, are shown in successive columns. These flows were chosen as
exemplar of the various different dynamic regimes. The flow at $Ro =
1.5$ is defined by low fluctuation levels of both the kinetic and
magnetic fields. The mean flow is decidedly quasigeostrophic, with
only the barest nub of a superrotary region near the equator of the
inner sphere. The $Ro = -1.0$ flow has a quasigeostrophic region near
the outer edge, but the dominant feature of both the mean flow and the
fluctuations is a large superrotary region extending into bulk flow
where the strongest of the kinetic and magnetic fluctuations live. The
$Ro = -2.0$ flow starts to return to quasigeostrophy, although the
bulk flow has not quite reached a state of z invariance. Here the
fluctuations of the kinetic energy live predominantly towards the
outer sphere, while the strongest magnetic fluctuations are
concentrated more towards the inner sphere. This is not particularly
surprising, as the applied magnetic field is 27 times stronger at the
inner sphere than the outer.

In section~\ref{sec:spattemp} we will see that the fluctuations at $Ro
= -1.0$ are characterized by persistent structures with a fairly well
defined spatial location, while the fluctuations at $Ro=-2.0$ are
characterized by intermittently formed structures that propagate
poleward from their point of origin. This means that the time averages
of Fig.~\ref{fig:spaceavg}b give a good indication of the shape of
the fluctuations being averaged over, while the time averages of
Fig.~\ref{fig:spaceavg}c smear the structures out over the region they
propagate through. As the fluctuations in the case of positive
differential rotation were significantly smaller in scale and in
amplitude than those for negative differential rotation, they will be
ignored in the following discussion.

\section{Spatio-temporal analysis}
\label{sec:spattemp}
\subsection{Modal flows}
\label{sec:pca}

\begin{figure}
  \includegraphics[width=0.5\linewidth]{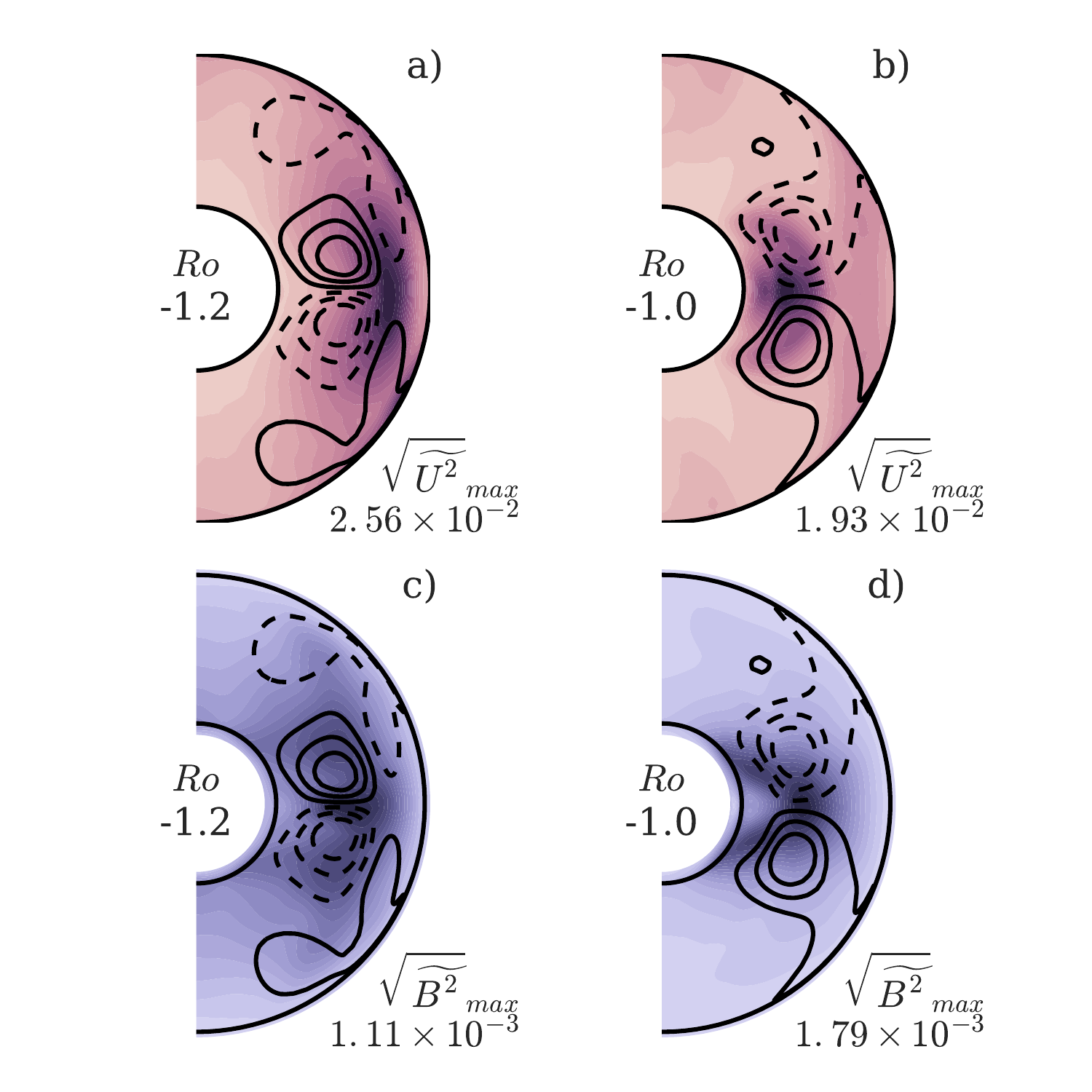}
  \caption{As in Fig.~\ref{fig:spaceavg}, but only for modal
    flows. The contour lines represent streamlines of poloidal
    circulation. Solid lines are counter-clockwise circulation; dashed
    lines are clockwise circulation. All of the energies are
    calculated for $\Lambda_b = 0.095$. \label{fig:modespaceavg}}
\end{figure}

The flows at $Ro = -1.2, -1.0$ revealed fluctuations living deep in
the bulk of the flow.  Figure~\ref{fig:modespaceavg} shows the time
and azimuthally averaged kinetic and magnetic fluctuations at $Ro =
-1.2, \Lambda_b = 0.095$ (a and c respectively) and for $Ro = -1.0,
\Lambda_b = 0.095$ (b and d). The streamlines of the mean meridional
flow are plotted over each map. Both cases include an outward
streaming jet originating at the equator of the inner sphere and an
inward streaming jet originating at the equator of the outer
sphere. The difference is that at $Ro = -1.2$ the outward jet
dominates the mean flow, extending to a radius of 0.82, while at $Ro =
-1.0$ the inward jet dominates from a radius of 0.40. In both cases
the oppositely directed jet is too weak to be seen in the streamlines.

\begin{figure}
  \includegraphics[width=0.8\linewidth]{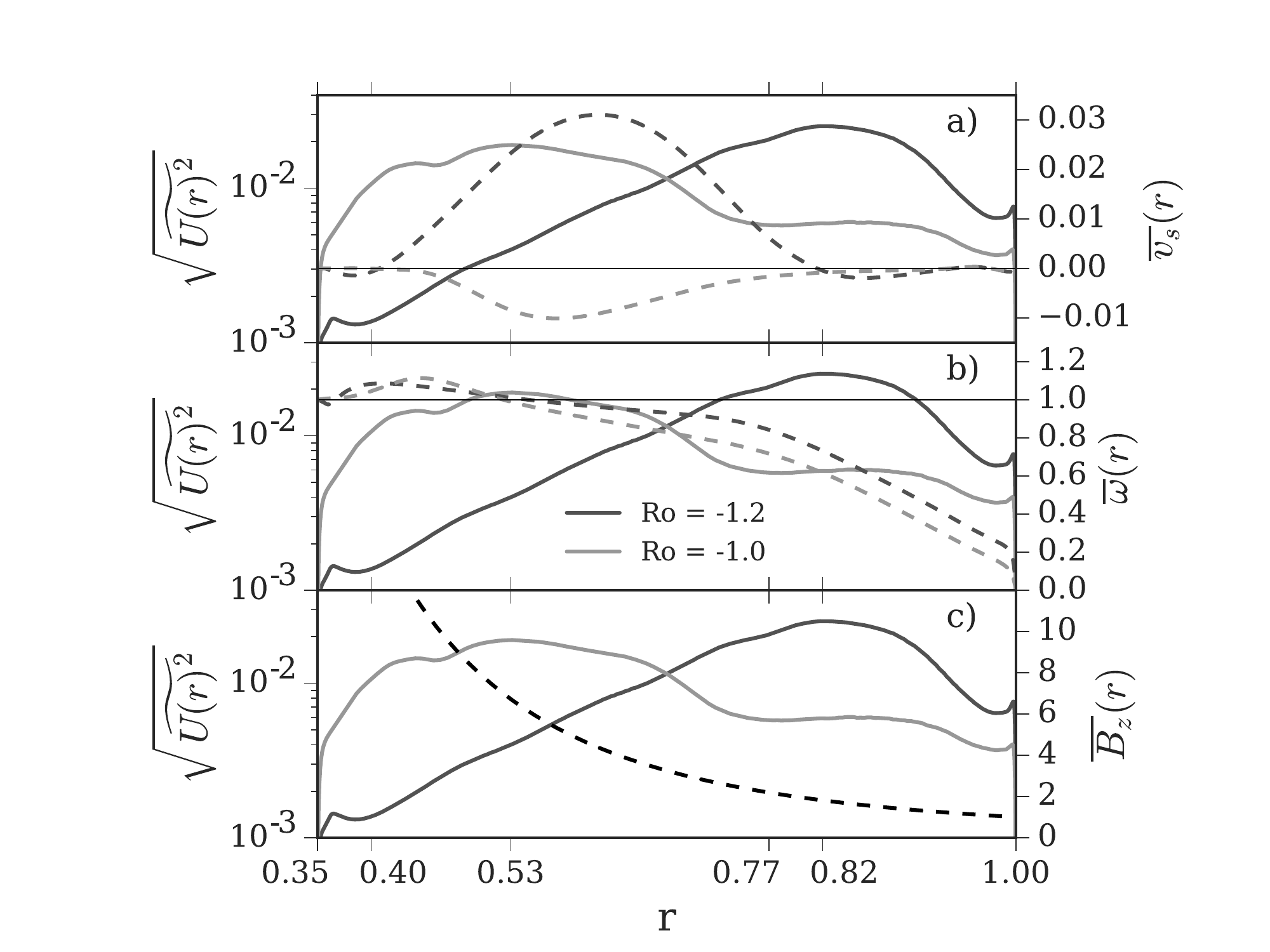}
  \caption{Kinetic fluctuation levels at the equator (solid lines) for
    Ro = -1.2 and Ro = -1.0 and $\Lambda_b = 0.095$ in log-scale (on the left)
    as a function of $r$.  The (a) {bulk flows' cylindrically radial velocity,} (b)
    {bulk flows' angular velocities,} and (c) {imposed vertical magnetic field
      in the equatorial plane} are plotted as dashed lines with the
    vertical scales on the right. The radii of the peak fluctuation
    levels are noted on the x-axis.
    \label{fig:fieldatinstab}}
\end{figure}

The kinetic fluctuations at $Ro=-1.2$ live predominantly near the
stagnation point, $\left(u_s=0\right)$; those at $Ro=-1.0$ live in
the jet itself $\left(\left|u_s\right|_{max}\right)$ as shown in
Fig.~\ref{fig:fieldatinstab}. The jets velocities are shown in
Fig.~\ref{fig:fieldatinstab}a. The azimuthal flows, shown in
Fig.~\ref{fig:fieldatinstab}b, and background magnetic fields, shown
in Fig.~\ref{fig:fieldatinstab}c, do not seem to determine whether a
location is stable or not, but their profiles are noted for a later use.

\begin{figure}
  \includegraphics[width=\linewidth]{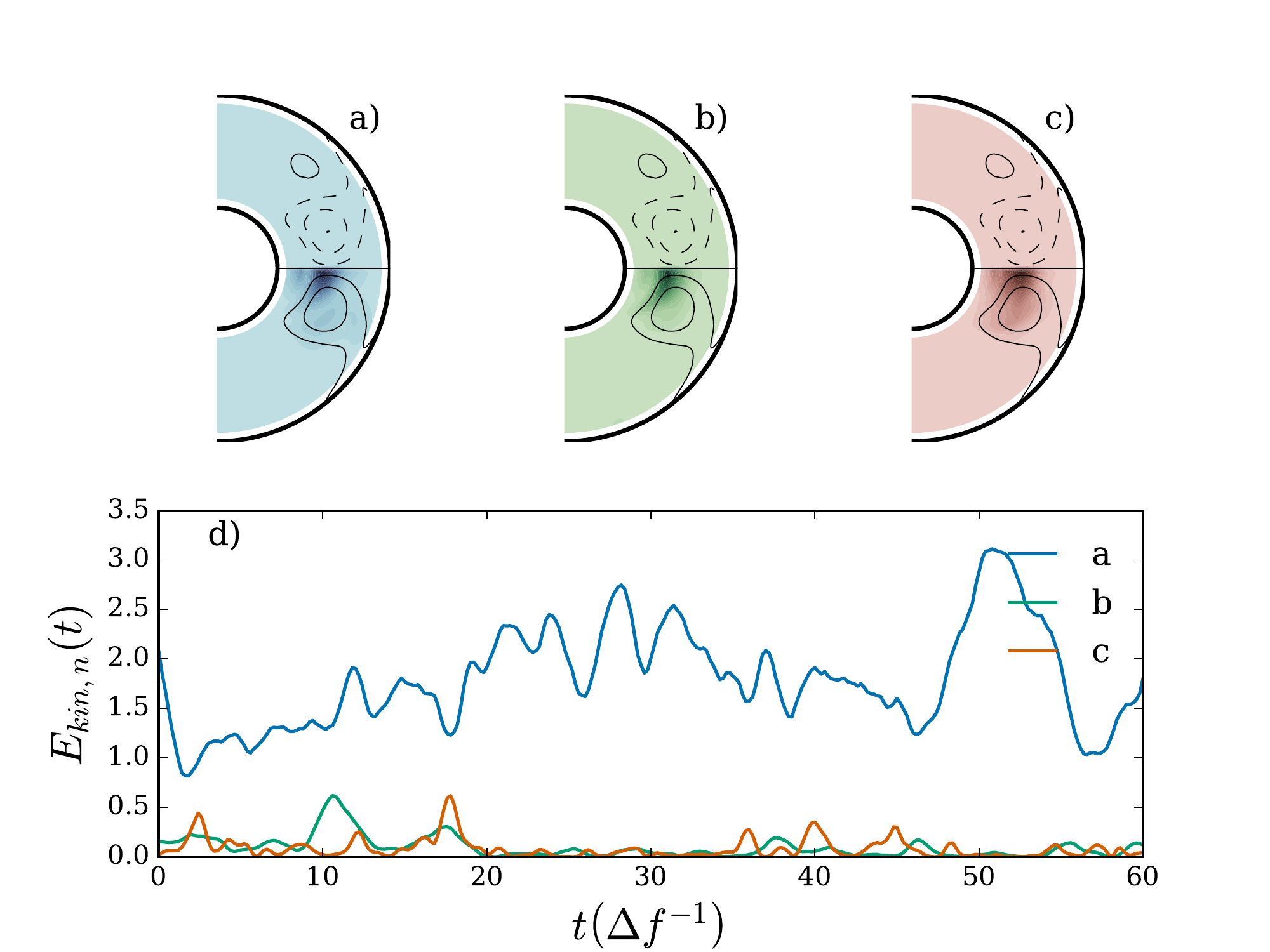}
  \caption{(a-c) Colormaps showing azimuthally averaged kinetic energy density of the
    first three singular modes of the $Ro = -1.0, \Lambda_b = 0.095$
    flow. The upper hemisphere shows the equatorially antisymmetric
    component of the mode; the lower hemisphere shows the equatorially
    symmetric component. (d) The energy in each mode (a-c) as a
    function of time, scaled against the mean energy in the kinetic
    fluctuations of the raw signal. Only the region $r \in [0.425,
      0.975]$ is included in the PCA. \label{fig:leftright-1}}
\end{figure}

\begin{figure}
  \includegraphics[width=\linewidth]{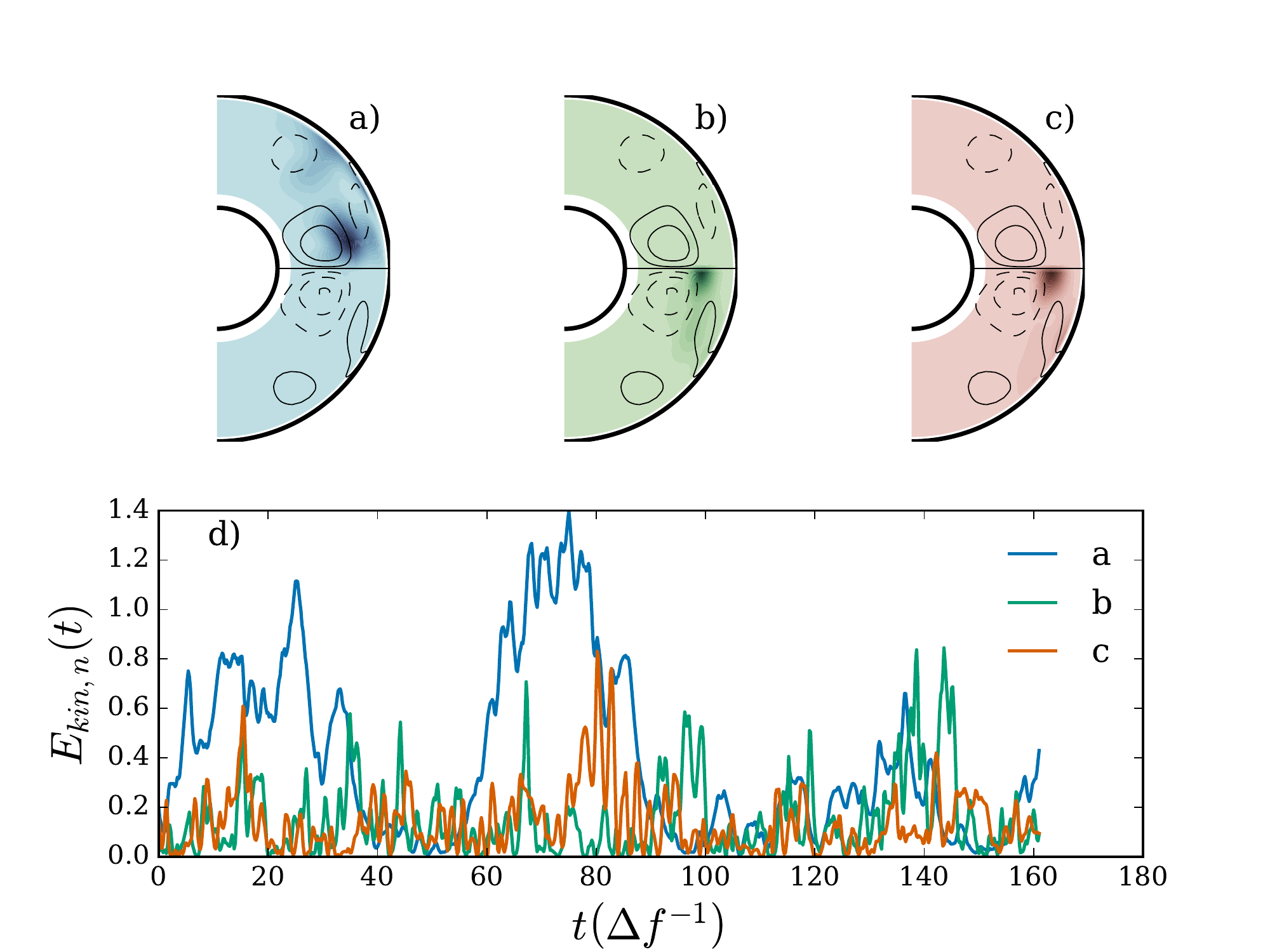}
  \caption{(a-c) Colormaps showing azimuthally  averaged kinetic energy density of the
    first three singular modes of the $Ro = -1.2, \Lambda_b = 0.095$
    flow. The upper hemisphere shows the equatorially antisymmetric
    component of the mode; the lower hemisphere shows the equatorially
    symmetric component. (d) The energy in each mode (a-c) as a
    function of time, scaled against the mean energy in the kinetic
    fluctuations of the raw signal. Only the region $r \in [0.425,
      0.975]$ is included in the PCA. \label{fig:leftright-2}}
\end{figure}

These fluctuations proved susceptible to a principle component
analysis (PCA, see Appendix~\ref{appendix:PCA}), where a
spatio-temporal signal is decomposed into individual spatial
structures that evolve independently of each other in time. The first
three of these structures for $Ro = -1.0, \Lambda_b = 0.095$, averaged
in azimuth and split into equatorially antisymmetric components in the
'northern' hemisphere and equatorially symmetric components in the
'southern' hemisphere, are shown in Fig.~\ref{fig:leftright-1} (a-c)
with their time series shown in Fig.~\ref{fig:leftright-1}d. The
streamlines of the meridional flow are plotted over the maps. All
three modes are very clearly symmetric about the equator, and the
first mode very clearly dominates the others over the time period
analyzed. The same decomposition is shown for $Ro = -1.2, \Lambda_b =
0.095$ in Fig.~\ref{fig:leftright-2}. The dominant mode (a) is
antisymmetric about the equator, but it arises in chaotic bursts for
these parameters.  (This mode is more clearly dominant at $\Lambda_b =
0.19$, see Fig.~\ref{fig:leftright-4} in Appendix~\ref{app:ModalFlows}).

\begin{figure*}
  \includegraphics[width=0.9\linewidth]{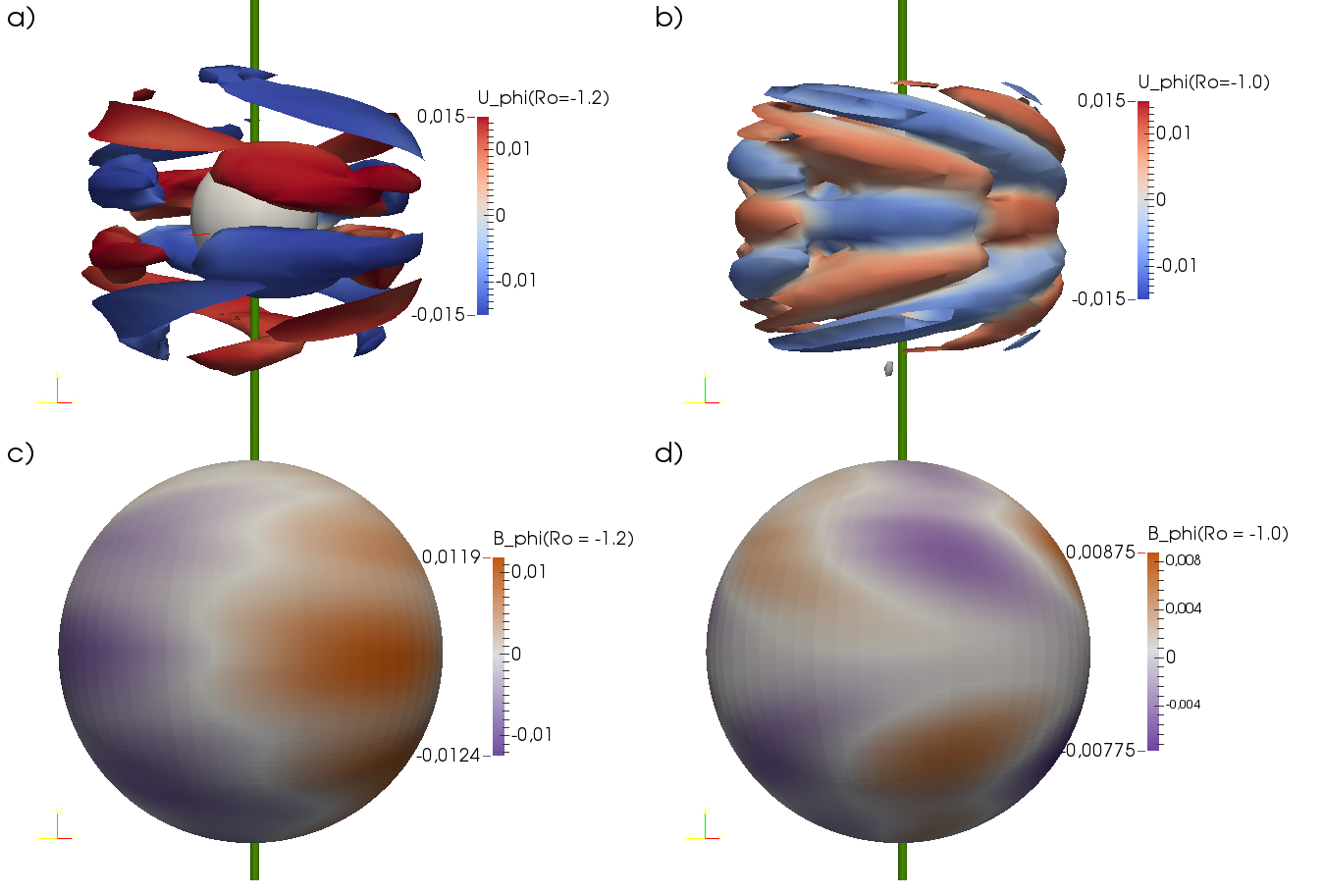}
  \caption{\texttt{Paraview} rendering of the velocity component (a)
    and the surface magnetic field (c) of the first principle component
    at $Ro = -1.2, \Lambda_b = 0.095$. Figures b) and d) are the same for $Ro
    = -1.0, \Lambda_b = 0.095$. The surfaces in a) and b) are isocontours of
    $\left|\uv\right|$, at $0.15 U_0$ and $0.10 U_0$ respectively,
    with the azimuthal velocity painted on the surface. The azimuthal
    magnetic field of the components (scaled against $B_o$). at the
    surface of the outer sphere are shown in c) and d). The inner
    sphere is visible in (a); the green cylinders represent the
    rotation axis.
    \label{fig:svdrender}}
\end{figure*}

Figure~\ref{fig:svdrender} shows 3-d renderings of the dominant
principal component, both the velocity field (a,b) and the surface
magnetic field associated with it (c,d) for $Ro = -1.2$ (a,c) and $Ro
= -1.0$ (b,d). The dominant mode is antisymmetric at $Ro = -1.2$ and has a predominant azimuthal mode number $m=2$, while it is
symmetric at $Ro = -1.0$ with $m=3$. It is worth noting that the background
magnetic field, shown in Fig.~\ref{fig:fieldatinstab}(c) is $3.5 \times$
larger at the $Ro = -1.0$ fluctuation peak than at the $Ro = -1.2$
peak.

This behavior connects the modes seen here with those found in other
simulations of MSC flow. For MSC flow without global rotation and a
vertical, homogeneous applied field, Hollerbach
\cite{Hollerbach.RSPA.2009} found equatorially antisymmetric
instabilities of the equatorial jet at low applied magnetic field and
equatorially symmetric instabilities of the stagnation point at an
applied field about $2.5 \times$ stronger (depending on the rotation
of the inner sphere). The magnetic fields in these simulations were
comparable to those performed here (equivalent to $10^{-3} < \Lambda <
5 \times 10^{-2} $, see
Appendix~\ref{app:SimulationResolution}). Continuing on this,
Gissinger {\it et al.} \cite{Gissinger.PRE.2011} explored a similar
system with a dipole magnetic field and global rotation (equivalent to
$Ro = 3.5, \;0.007 \leq \Lambda_o \leq 2.8$ by our
definitions \footnote{The ratio of rotation rates, $\Lambda$s, and
  Reynolds numbers previously reported in \cite{Gissinger.PRE.2011}
  contained typos. Private Communication}) and found a similar set of
instabilities changing from antisymmetric to symmetric about the
equator as the applied magnetic field was increased. The modes found
here are novel in that the jet's instability is equatorially symmetric and the
return flow instability is antisymmetric. This suggests that the
actual stability of the background flows in all of these cases is
agnostic to the magnetic field entirely, but that the symmetry is
determined by the strength of the vertical field where the instability
arises. It is worth noting that the other principle components of
$Ro=-1.2$ resemble the principle component of $Ro=-1.0$ in shape and
symmetry.
\subsection{Filamentary flow}
\label{sec:filaments}

\begin{figure}
  \includegraphics[width=0.45\linewidth]{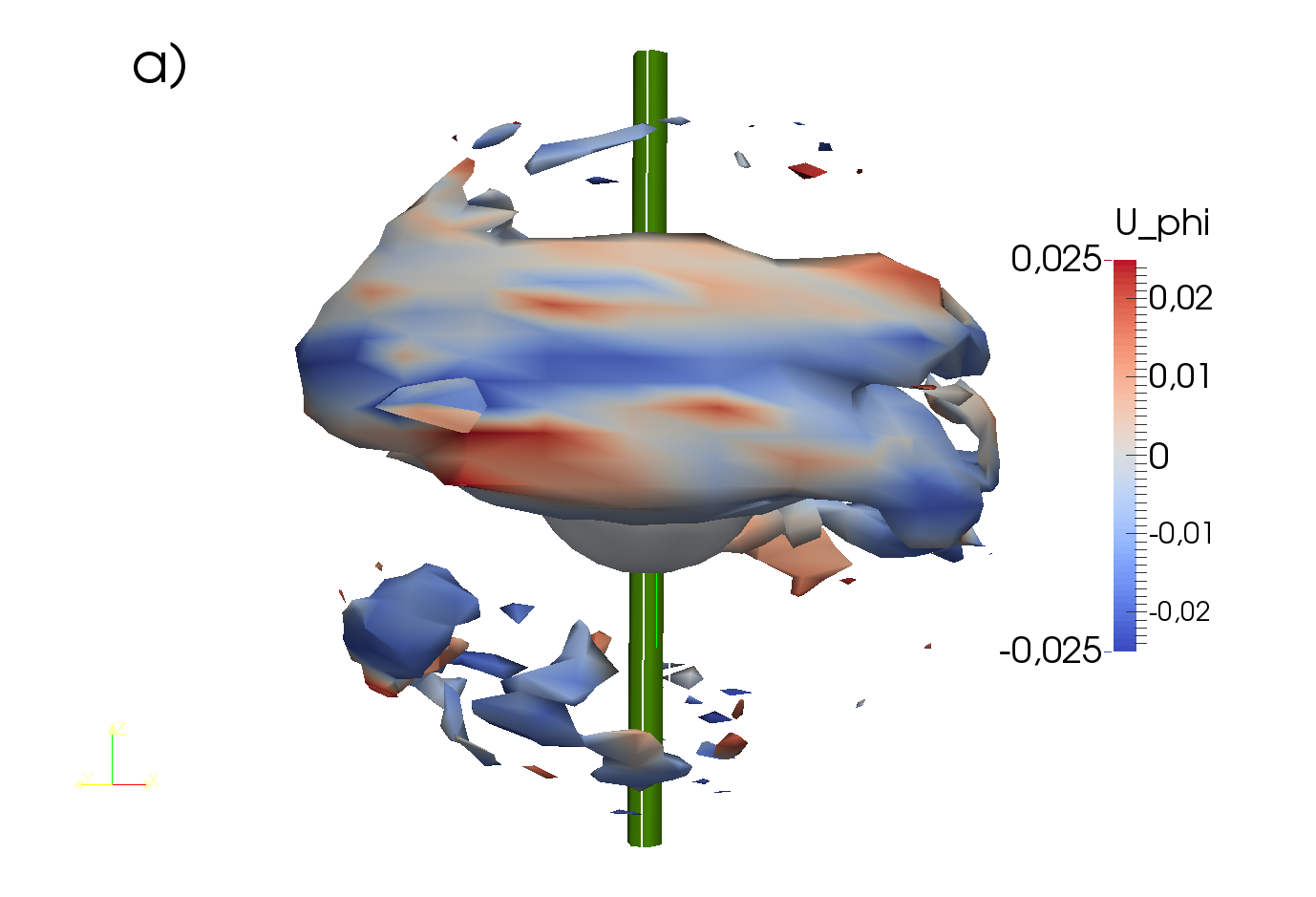}
  \includegraphics[width=0.45\linewidth]{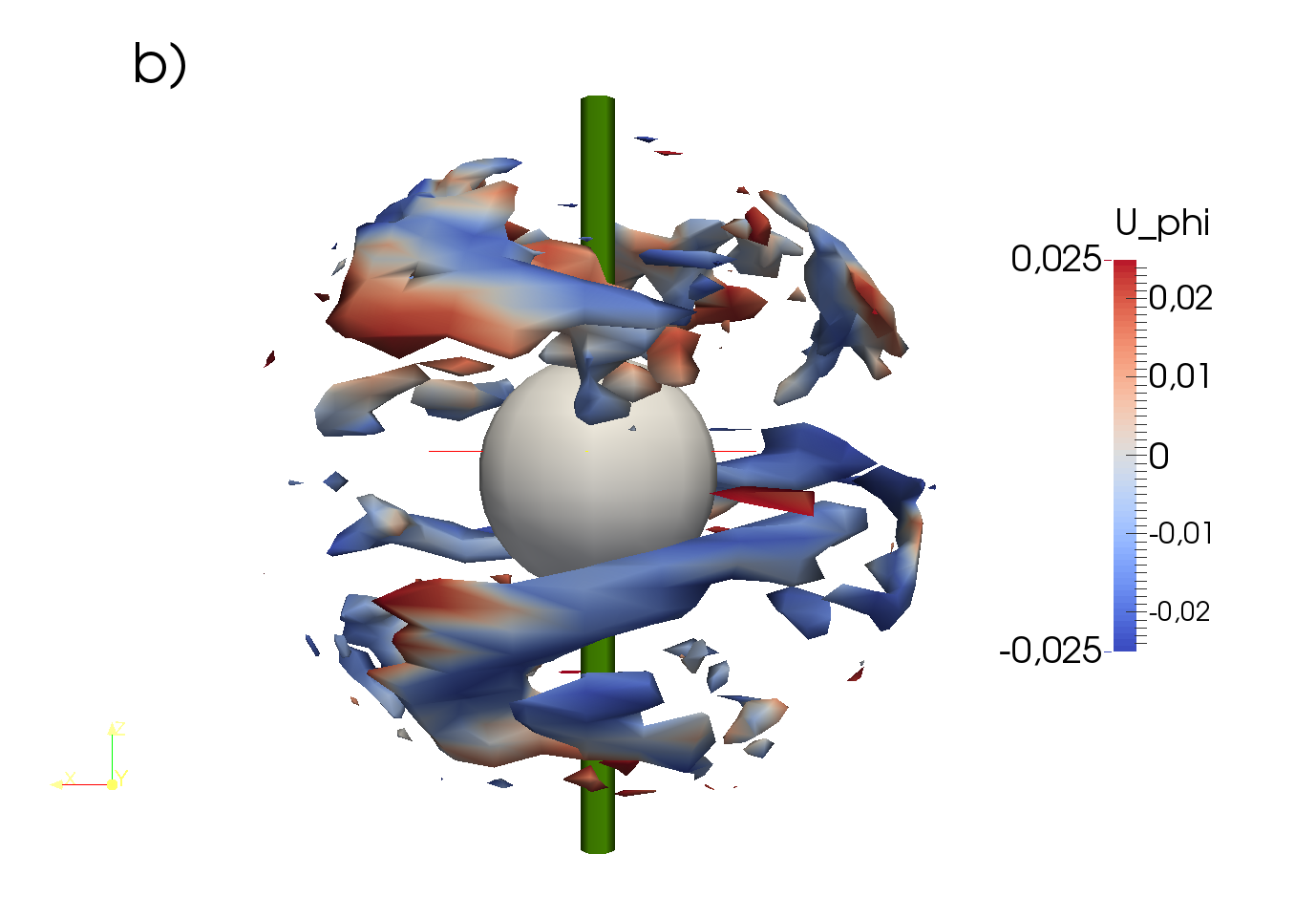} \\
  \includegraphics[width=0.45\linewidth]{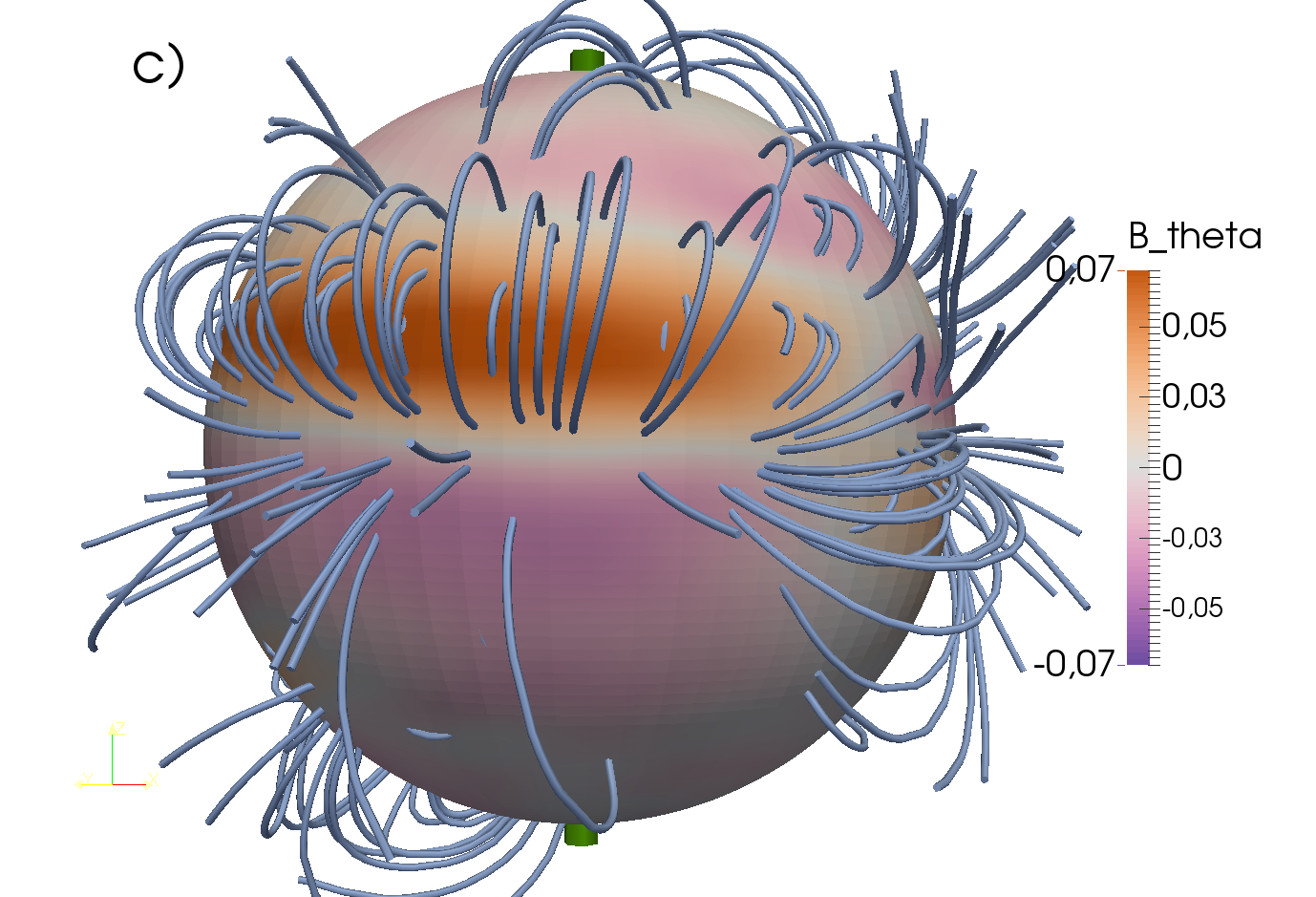}
  \includegraphics[width=0.45\linewidth]{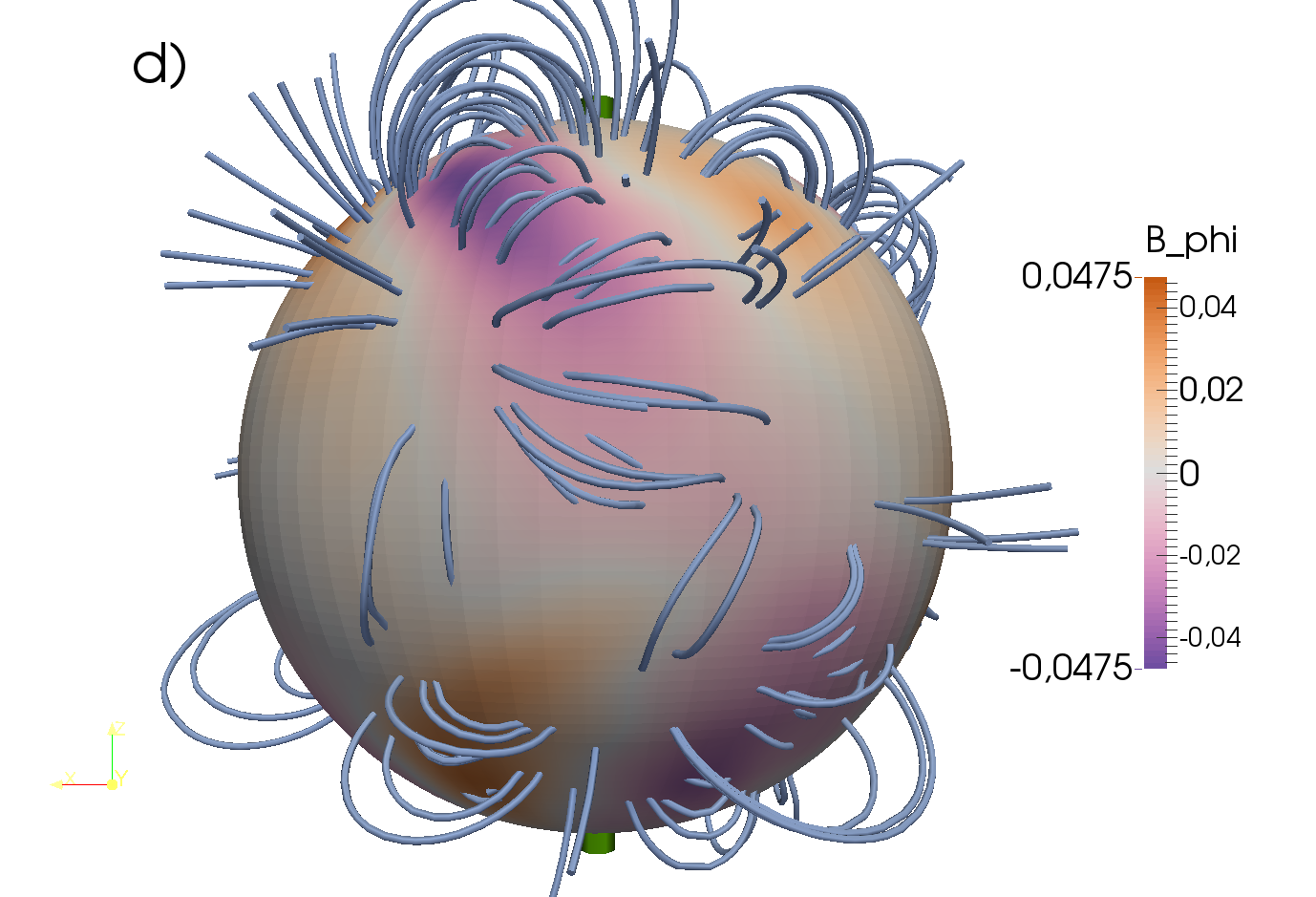}
  \caption{\texttt{Paraview} \cite{PARAVIEW} rendering of two
    individual snapshots of the flow at $Ro=-2.0$ and $\Lambda_b =
    0.095$. The renderings show (a) an isosurface of
    $\left|\widetilde{\mathbf{v}}\right|$ corresponding to 0.025 $U_0$
    and (b) the same contour $16 \Delta f^{-1}$ later. The isosurfaces
    are painted with the azimuthal velocity. The surface magnetic
    fields, scaled against $B_o$ of these snapshots are shown in (c)
    and (d) respectively, along with the magnetic field lines
    originating at the isocontours. The latitudinal field is shown for
    (c), the azimuthal field for (d). The green cylinders represent
    the rotation axis. \label{fig:filament}}
\end{figure}

When the differential rotation is increased beyond $Ro = -1.2$, the
fluctuations move out of the bulk flow and take up residence along the
outer sphere. Here they form long, thin filaments, at predominantly
m=1, that are carried towards the pole by the meridional
circulation. At these higher latitudes, they split up along the
azimuthal coordinate. Figure~\ref{fig:filament} shows (a) a 3-d
rendering of an isocontour of kinetic energy and (c) the latitudinal
magnetic field at the surface and the magnetic field lines originating
on the kinetic isocontours and extending out of the sphere. The same
is shown at a time 16 differential rotation periods later, as the
filament is breaking up, in (b \& d), though the surface field is
azimuthal in (d). The magnetic field lines extending from the kinetic
structures into the vacuum are also shown in (c \& d). These snapshots
were taken from a simulation at $Ro=-2.0$ and $\Lambda_b = 0.095$

\begin{figure}
  \includegraphics[width=0.8\linewidth]{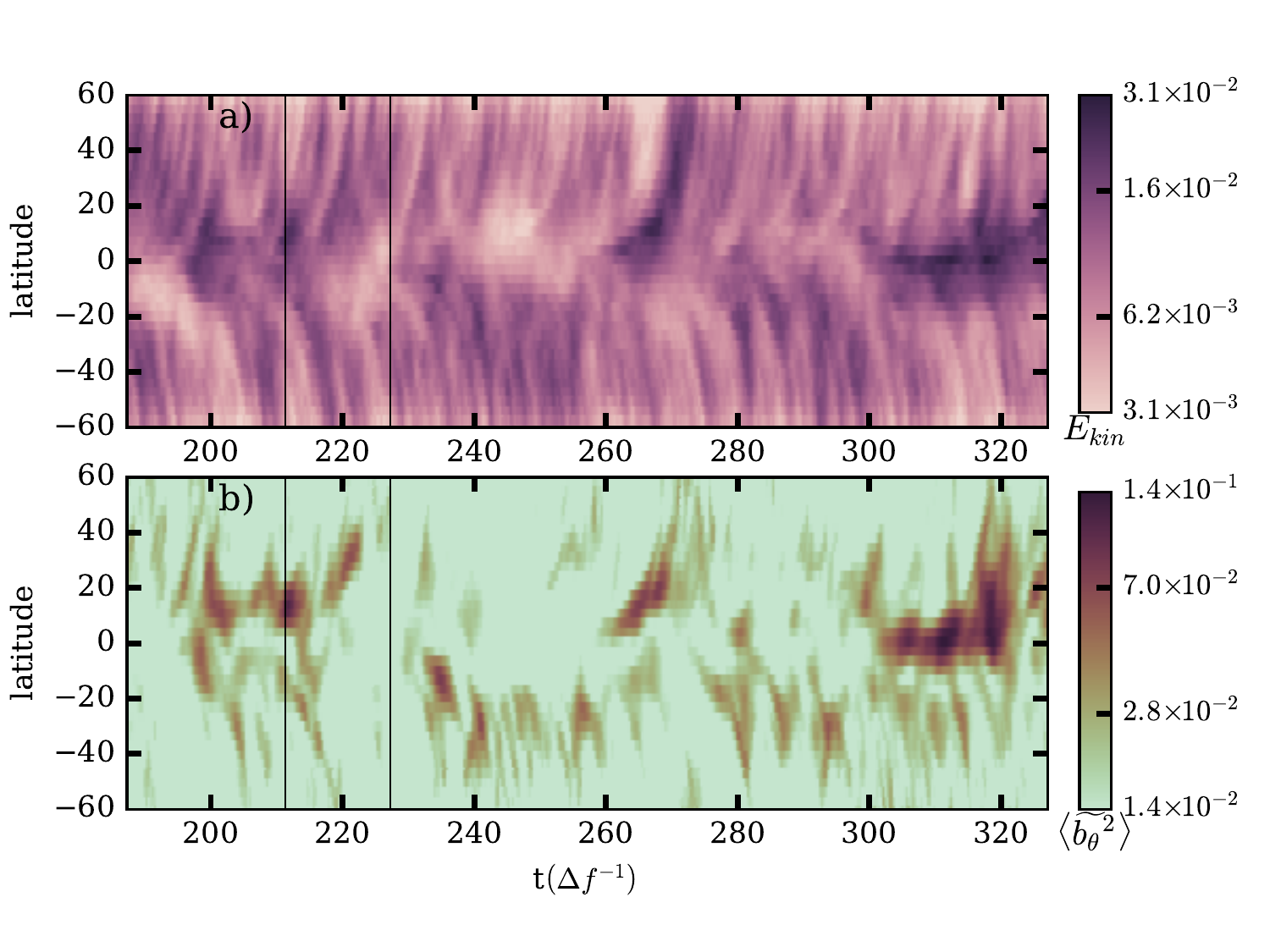}
  \caption{a) Cylindrically Integrated kinetic energy vs latitude and
    time. b) Azimuthally integrated $\widetilde{b_\theta}^2$ at the
    surface (scaled against $\left.B_o\right)$ vs latitude. Filamentary regime at
    $Ro=-2.0$ and $\Lambda_b = 0.095$.  The vertical lines indicate the
    snapshots the renderings of Fig.~\ref{fig:filament} are taken
    from. \label{fig:integralu_surfaceb}}
\end{figure}

The surface magnetic field seems to be a good indicator of the flow
underneath. Figure~\ref{fig:integralu_surfaceb}a shows the
cylindrically integrated kinetic energy,
\begin{equation}
  E_{kin}(z, t) = \int \limits_0^{2\pi}d\phi \int \limits_0^{\sqrt{1-z^2}}ds\;
  \mathbf{\widetilde{u} \cdot \widetilde{u}}(s, z, \phi, t),
\end{equation}
\noindent where $s$ is the cylindrical radius $s = r\sin\theta$ and
$z$ is the cylindrical height $z=r\cos\theta$. The latitudinal field
at the surface, averaged in $\phi$ is shown in
Fig.~\ref{fig:integralu_surfaceb}b. The times Fig.~\ref{fig:filament}
are taken from are indicated. The latitudinal field component
corresponds well with the kinetic fluctuations near the equator (below
$\sim 40^\circ$). The azimuthal field (not shown) has a stronger
correspondence towards the poles. As seen in Fig.~\ref{fig:filament},
the kinetic fluctuations are elongated more in $\phi$ towards the
equator and more in $\theta$ towards the pole; the induced magnetic
field seems strongest in the direction perpendicular to the
elongation. This will be shown more in Section~\ref{sec:time}.

Previous simulations of the experiment with a stationary outer sphere
also turned up two classes of fluctuations: a centripetal jet
instability near the equator, and a B\"odewalt type instability
originating in the upper latitudes (above $\sim$40$^\circ$) both
propagating poleward \cite{Figueroa.JFM.2013}. The B\"odewalt type
instability is likely not present here, as the angular velocity
profile at the outer sphere's pole (\cite{Figueroa.JFM.2013} Fig.~3)
is replaced by an Ekman type boundary layer when the outer sphere
rotates. The centripetal jet instabilities demonstrated some similar
behavior to the filaments seen here. Specifically, structures
generated near the outer sphere's equator are swept poleward by the
meridional flow.

\section{Linear instabilities of the mean fields}
\label{sec:linear}
What is the origin of the modes observed in the various regimes? We use the time- and phi-averaged fields from the non-linear simulations as base flow $\bar{\mathbf{U}}$ and magnetic field $\bar{\mathbf{B}}$ for linear stability tests.
Note that for these base fields, we keep only the dominant symmetry: $\bar{\mathbf{U}}$ is symmetric while $\bar{\mathbf{B}}$ is anti-symmetric with respect to the equatorial plane.
We time-step the linearized MHD equations for the velocity and magnetic field perturbations $\mathbf{u}$ and $\mathbf{b}$ using the linearized \texttt{xshells} code \cite{vidal2018}, waiting for the growth-rate to stabilize.
We distinguish hydrodynamic stability ($\bar{\mathbf{B}}=0$, hence no Lorentz force), purely magnetic stability ($\bar{\mathbf{U}}=0$, but $\nabla \times \bar{\mathbf{B}} \neq 0$) from magnetohydrodynamic (MHD) stability (using both $\bar{\mathbf{B}}$ and $\bar{\mathbf{U}}$).
By comparing these three cases, we can determine if the MHD instability is merely a hydrodynamic instability, or if the magnetic field plays a more fundamental role.
Results are summarized in table \ref{tab:stab}.
First, the mean magnetic configurations alone ($\bar{\mathbf{U}}=0$) are found always stable at the field strengths of this study.
Then, we find that the growth rate of the MHD perturbation is always smaller than the hydrodynamic one.
These two observations suggest that the mean magnetic field and associated electric current have no driving effect but rather a damping effect on instabilities driven by the mean velocity field.

\begin{figure}
  \includegraphics[width=0.22\linewidth]{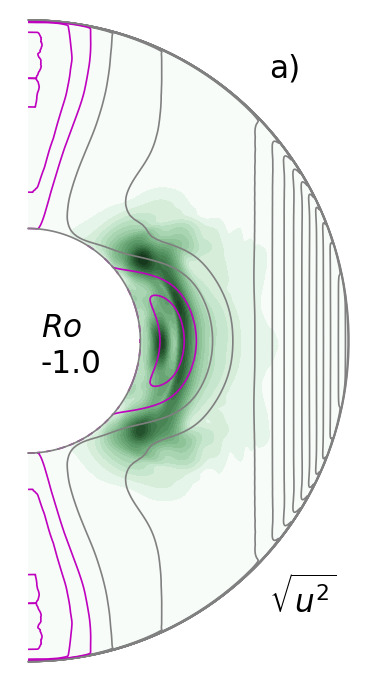}
  \hspace{0.1\linewidth}
  \includegraphics[width=0.22\linewidth]{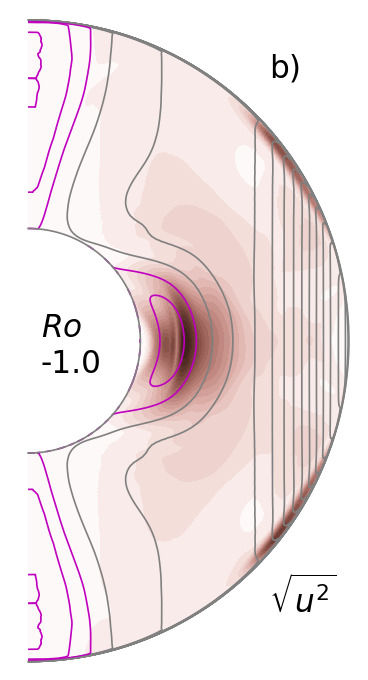}
  \includegraphics[width=0.22\linewidth]{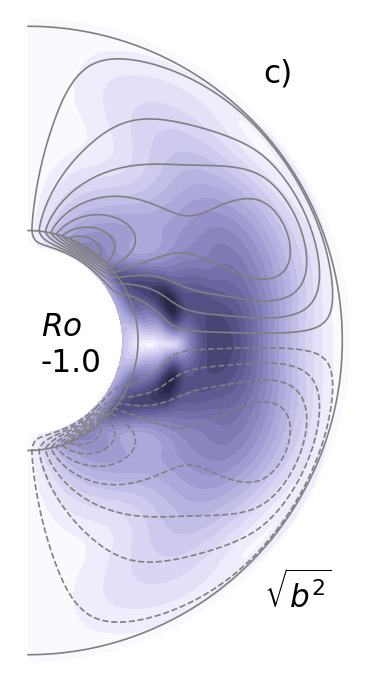} \\
  \includegraphics[width=0.22\linewidth]{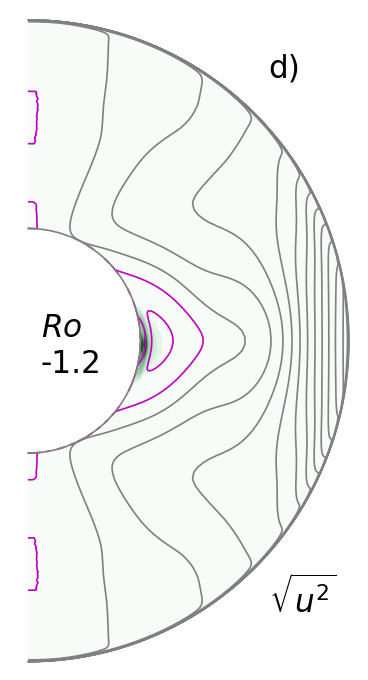}
  \hspace{0.1\linewidth}
  \includegraphics[width=0.22\linewidth]{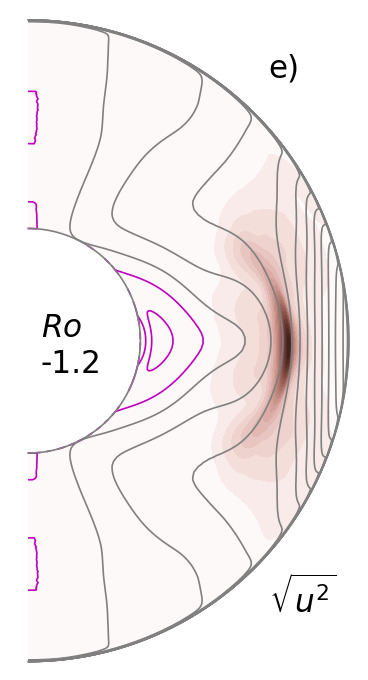}
  \includegraphics[width=0.22\linewidth]{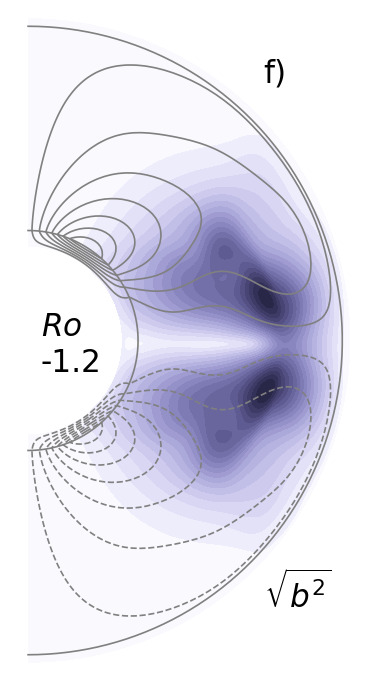}
  \caption{Colormaps of the intensity of the most unstable mode (white is zero). Top row (a,b,c): $Ro=-1.00$ and $\Lambda_b = 0.095$. Bottom row (d,e,f): $Ro=-1.20$ and $\Lambda_b = 0.095$. Left (a,d) shows the hydrodynamic instability while  middle (b,e) and right (c,f) represent respectively the velocity and magnetic intensity of the magnetohydrodynamic instability. The contours in (a,b,d,e) are iso-lines of the angular velocity of the mean flow $\bar{\mathbf{U}}_\phi/s$ with magenta for super-rotation. The contours in (c,f) are iso-lines of the the toroidal mean magnetic field $\bar{\mathbf{B}}_\phi$ (dashed for negative).
All velocity eigenmodes are equatorially symmetric while the magnetic ones are anti-symmetric.
  \label{fig:stab_merid}}
\end{figure}

Looking at individual cases, we see that for $Ro=-2.0$, the mean flow is subject to a strong hydrodynamic instability of the outer boundary layer, where the magnetic field is weaker. This happens for the two base flows obtained with different field strength (as measured by $\Lambda_b$).
For $Ro=-1.0$ and $Ro=-1.2$, the intensity of the perturbations have been reported in figure \ref{fig:stab_merid}.
For $Ro=-1.2$, depending on the field strength $\Lambda_b$, the mean flow may be subject to a similar hydrodynamic instability of the outer boundary layer (but for a lower wavenumber) or to an instability localized at the equator of the inner-sphere (Fig. \ref{fig:stab_merid}d).
In both cases, this hydrodynamic instability does not survive the additional damping due to the magnetic field, which favors a bulk instability of low wave number ($m=4$, Fig. \ref{fig:stab_merid}e,f).
For $Ro=-1.0$, both hydrodynamic and MHD instabilities occur away from the boundary layers. They are similar, but the MHD instability seems to be slightly pushed away from the strong field inner-core to a region were the magnetic field is lower (compare Fig. \ref{fig:stab_merid}a and b).
Note that for all cases investigated, the MHD perturbations show little dependence upon $\Lambda_b$.

\begin{table}
\begin{tabular}{ll|rcl|rcl}
\toprule
 & & \multicolumn{3}{c|}{hydrodynamic instability} & \multicolumn{3}{c}{MHD instability} \\
$Ro$  & $\Lambda_b$ &  $m$ & $\sigma$ & localization & $m$ & $\sigma$ & localization  \\
\hline
-2.00 & 0.190 & 22 & 1.59 & outer boundary layer & 22 & 1.55 & outer boundary  layer  \\
      & 0.095 & 22 & 1.30 & outer boundary layer & 22 & 1.29 & outer boundary  layer \\
-1.20 & 0.190 &  5 & 0.264 & outer boundary & 4 & 0.198 & like fig. \ref{fig:stab_merid}e,f \\
      & 0.095 &  18 & 0.2961 & see fig. \ref{fig:stab_merid}d & 4  & 0.2235 & see fig. \ref{fig:stab_merid}e,f \\
-1.00 & 0.190 &  3 & 0.1931 & like fig. \ref{fig:stab_merid}a & 2 & 0.0483 & like fig. \ref{fig:stab_merid}b,c  \\
      & 0.095 &  3 & 0.1828 & see fig. \ref{fig:stab_merid}a & 2 & 0.0569 & see fig. \ref{fig:stab_merid}b,c \\
\bottomrule
\end{tabular}
  \caption{Instability of the mean flow of several runs (see table \ref{tab:resolution} for the corresponding non-linear runs). $m$ is the unstable azimuthal wave number, $\sigma$ is the non-dimensional growth rate, and `localization' indicates the localization of the instability in the meridional plane.
  \label{tab:stab}}
\end{table}

We can now compare these instabilities with the fully nonlinear simulations.
First, at $Ro=-1.0$ and $\Lambda_b=0.095$ the PCA analysis shows a mode dominated by an $m=3$ symmetry (Fig. \ref{fig:leftright-1} and \ref{fig:svdrender}b), while the most linearly-unstable mode has $m=2$. The detailed structure differ (not shown), but the same equatorial symmetry is recovered.

We now turn to $Ro=-1.2$ and $\Lambda_b=0.095$ for which the PCA analysis shows a mode dominated by $m=2$ and anti-symmetric with respect to the equator (Fig. \ref{fig:leftright-2} and \ref{fig:svdrender}a).
Instead, the linearly-unstable mode has $m=4$ and is symmetric (not shown).
Comparing figure \ref{fig:leftright-2} with figure \ref{fig:stab_merid}e, we see that the most unstable MHD mode resembles the second and third PCA (labeled b and c respectively) but not the first one which has opposite symmetry.
The PCA analysis thus suggests an interplay between two MHD instabilities, the most unstable (symmetric) and probably the second most unstable (anti-symmetric).

Finally, for $Ro=-2$, the most unstable mode is located at the outer boundary (not shown) and we expect it to continuously generate turbulent flow that fills the bulk and leads to the filamentary regime.

All these observations support that instabilities are triggered by the mean flow, while the applied magnetic field selects, damps and reshapes them, keeping in mind that the mean flow itself is strongly affected by the presence of the applied magnetic field.
Interestingly, large-scale coherent structures emerging from hydrodynamical turbulence have also been linked to linear stability of their base flow \cite{herbert2014,brethouwer2014}, while exhibiting bursts as seen in figure \ref{fig:leftright-2}.
It is thus likely that in our system, linear instabilities growing on the mean flow lead to the persistent or intermittent structures observed in the various turbulent flow regimes described here.

\section{Simulated Probes}
\label{sec:time}
The trade-off between experimental and numerical physics is that it is
easy to get time series of experimental data, but difficult to extend
that data outside a small domain that is amenable to direct
measurement, while it is easy to see all of the structures within a
simulation at the cost of large amounts of computing time. The longest
simulations presented here correspond to 10-20 s of runtime on \dts.
Given the full spatial data included in the simulation outputs, it is
easy to produce simulated probes and associate these signals to
specific structures. In this section we calculate the magnetic field
at the outer surface on a single meridian over from snapshots taken at
a rate of 50 per differential rotation period. All reported values for
magnetic measurements are scaled against the intensity of the magnetic field
at the outer sphere's equator $B_o$ for ease of comparison between the
simulations and the experiment (where $B_o$ is 7.4 mT). All of these
simulations were carried out at $\Lambda_b = 0.095$, equivalent to an
outer sphere rotation frequency of 10 Hz.
The results can directly be compared to the experimental examples of figure~\ref{fig:experiments}.

\subsection{Modal Flows}

\begin{centering}
  \begin{figure}
    \includegraphics[width=0.6\linewidth]{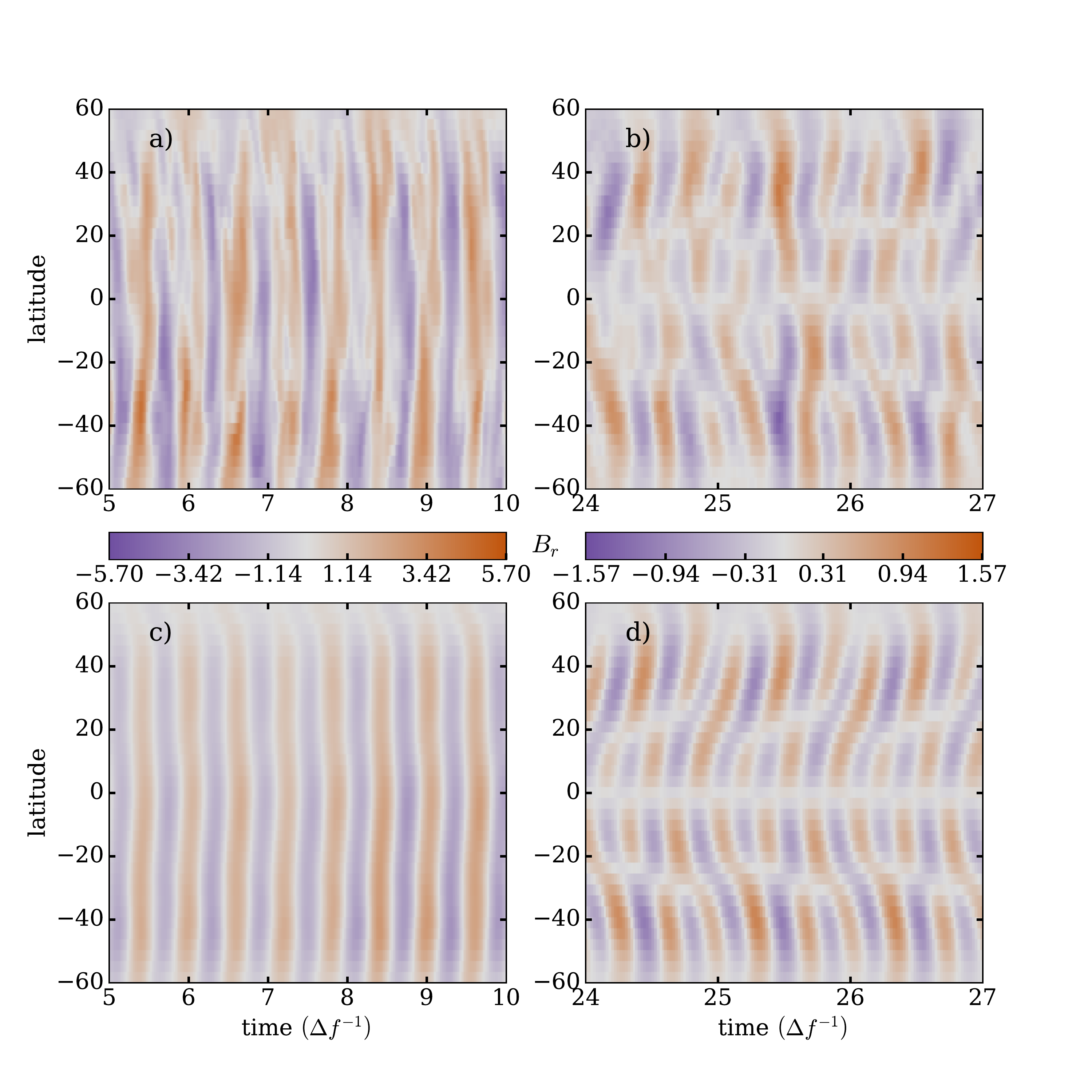}
    \caption{Radial component of the surface magnetic field (as a
      percentage of $B_o$ at a single meridian on the outer sphere vs
      time at a) Ro = -1.2 and b) Ro = -1.0 at $\Lambda_b = 0.095$. The
      signal from the dominant principle component is shown below in
      c) and d) respectively. \label{fig:surfpca}}
  \end{figure}
\end{centering}

The kinetic structures of Fig.~\ref{fig:svdrender} have specific
azimuthal orders (m) and equatorial symmetries and are rotating
relative to the outer sphere. Thus the magnetic field at the surface
of the sphere should behave entirely as a function of the shape and
rotation of the underlying structure. Figure~\ref{fig:surfpca} shows
the magnetic field at the surface for (a) Ro = -1.2 and (b) Ro = -1.0,
as well as the measurable field associated with the dominant principle
component for each (c and d respectively). As in
Fig.~\ref{fig:svdrender}, the magnetic structures at Ro = -1.2 are
vertical, and sweep past the probe with a frequency of $1.7\Df$
(rather, an m=2 structure rotating with a frequency of $0.85
\Df$). The magnetic structures at Ro = -1.0 are helical and sweep past
the probe with a frequency of $2.9\Df$ (rather an m=3 structure
rotating at $0.96\Df$). In both cases, the full signal strongly
resembles that of the dominant mode with additional signals on top.
This is not preordained, as the PCA was computed over the entire
volumetric dataset, and emphasizes again that the surface measurements
are strongly connected to the global dynamics.  The rotation rates are
related to the mean flow of the background at the location of the
structure; Fig.~\ref{fig:fieldatinstab} indicates the locations where
the bulk flow in the equator rotates at the rate of the structure ($r
= 0.53$ for $Ro = -1.0$, and $r = 0.77$ for $Ro = -1.2$).

\subsection{Filamentary Flows}

\begin{centering}
  \begin{figure*}
    \includegraphics[width=\linewidth]{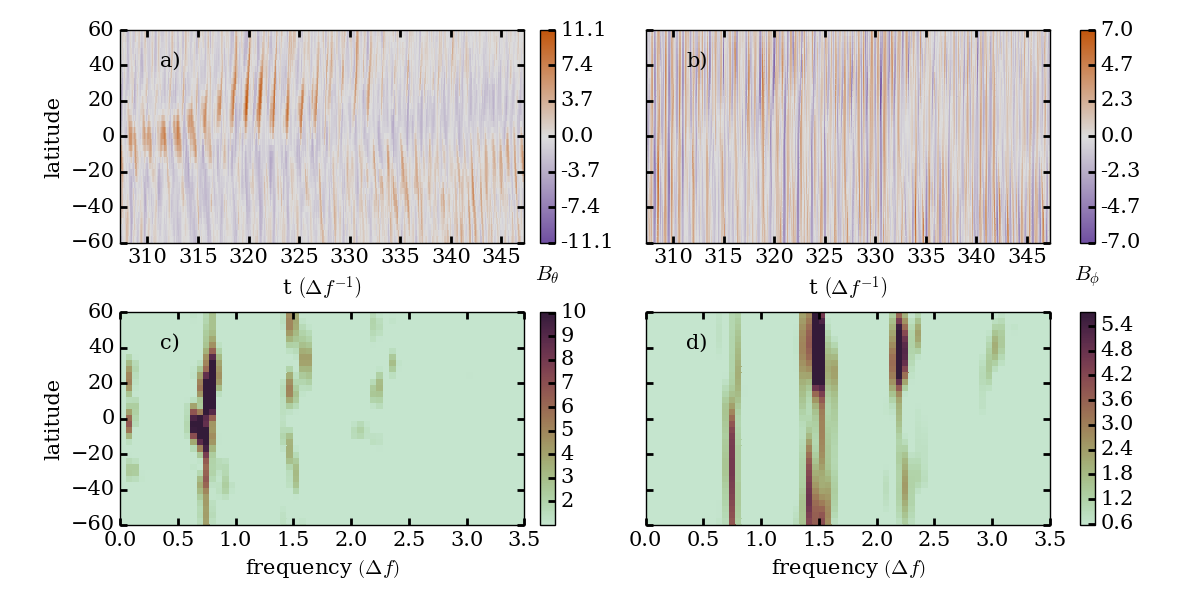}
    \caption{The surface magnetic field, as a percentage of $B_o$ at a
      single meridian on the outer sphere vs time at $Ro$ = -2.0 and
      $\Lambda_b = 0.095$, broken up into (a) latitudinal and (b)
      longitudinal components. The second row (c \& d respectively)
      shows the power spectral density of the magnetic signal at each
      latitude on that meridian. The color scale is truncated at one
      half the maximum spectral level to help highlight the higher
      harmonics. \label{fig:surffil}}
  \end{figure*}
\end{centering}

The filaments of Section~\ref{sec:filaments} have broader azimuthal
extent than the modes of the previous discussion. They are not easily
divided into single azimuthal modes numbers, but the volume renderings
of Fig.~\ref{fig:filament} show an apparent combination of m=1 and
m=2, depending on the latitude. The latitudinal and azimuthal surface
fields of Fig.~\ref{fig:surffil} a \& b (respectively), and their
spectra in Fig.~\ref{fig:surffil}c \& d provided some
corroboration. The latitudinal field has the strongest spectral energy
near the equator at a frequency of 0.75$\Delta f$, and noticeable
peaks near 1.5 and 2.25 $\Delta f$ (that is 2 or 3 times the
fundamental) at higher latitudes. The azimuthal field at the surface
is much stronger at higher latitudes and at the higher frequencies.
Note also the low frequency modulation at frequency $\simeq 0.1\Delta
f$, visible on the $B_\theta$ probes and their spectra, especially
around the equator and 20 to 30$^\circ$ degrees of latitude, most
certainly related to the poleward sweeping of the filaments (see
\ref{sec:filaments}).

\section{Conclusions}
\label{sec:conclusions}

We have carried out a set of magnetized spherical Couette (MSC) flow
simulations at nondimensional parameters relevant to the \dts
machine. These simulations have found a set of different dynamic
regimes. For positive (and weakly negative, $Ro \geq -0.5$)
differential rotation the flows were characterized by low fluctuation
levels on top of quasigeostrophic base flows. At stronger, negative
differential rotation, the flow consists of large scale internal
structures, associated with magnetic signatures at the surface.  At
moderate, negative differential rotation, $Ro$=-1.0 and -1.2, these
structures are persistent, with a specific azimuthal order, rotating
with the bulk flow. At stronger differential rotation, $Ro$=-2.0,
these structures are intermittent (with lifespan around 10
differential rotation periods), with large azimuthal extent, that
propagate poleward from their point of origin.

Given that simulations of MSC flow contain the full structures of the
kinetic and magnetic fields, it is easy to draw relations between the
one and the other. What is more difficult is taking the surface
magnetic fields alone, {\it i.e} the diagnostic which is most readily
available to the physical \dts device, and inferring the structure
of the velocity field that induces it. 
Nevertheless, some broad inferences can 
be made from the surface fields and their spectra. The shift from
persistent structures to intermittent filaments corresponds to a shift
from a single dominant frequency in magnetic field measurements to
multiple bands depending on latitude and direction of the
magnetic field.

The persistent structures have precedence in other simulations of
MSCF, including those carried out with no global rotation
\cite{Hollerbach.RSPA.2009} and those carried out with positive
differential rotation \cite{Gissinger.PRE.2011}. (That our modes occur
at different values of $\left|Ro\right|$ is not so
surprising. Positive and negative differential rotation are not the
same, but they can demonstrate similar dynamics with offset
$\left|Ro\right|$ \cite{Zimmerman.JGR.2014}). In all cases these
structures arise from hydrodynamic instabilities of the meridional
circulation (either in the equatorial jets or near the stagnation
points).  What is novel about the structures here is that the
equatorial symmetry of these instabilities is reversed from those of
\cite{Hollerbach.RSPA.2009,Gissinger.PRE.2011}.
We attribute this to the strength of the background magnetic field.
We have shown that, although the magnetic field determines the shape of the mean flow in all cases, these two dimensional flows are stable or unstable to three
dimensional hydrodynamic instabilities independent of the magnetic
field.
The effect of the magnetic field is then to select and reshape the instability.
By doing this, an instability with symmetry opposite to the expected hydrodynamic instability can be observed, depending on the magnetic field strength.
The symmetric instability at the stagnation point ($Ro=-1.2$) is compatible with an oscillation of that point within the plane of the equator as the
strength of the colliding jets oscillate. 
In contrast, the antisymmetric instability of the inward jet ($Ro=-1.0$) is reminiscent of the instability that deflects the jet up or down
\cite{DumasThesis,Hollerbach2006,Wicht.JFM.2013}.

The intermittent filaments seen here in the most turbulent cases share
some features with the inward jets emerging from the outer sphere and
drifting poleward in simulations of the DTS device without outer
rotation \cite{Figueroa.JFM.2013}.  However, the filaments evidenced
in this work appear to last significantly longer.

We emphasize that all the coherent structures seen in our simulations appear on top of a fully turbulent state.
Such a phenomenon has also been documented in hydrodynamical turbulence \cite{herbert2014,brethouwer2014}, where the instabilities of the mean flow play an important role too.

Our simulations have also shown that the surface measurements of the
magnetic fields are significantly affected by flow dynamics deep in
the shell.  This means that the many measurements that will be
available on the \dts machine will likely help to evidence the flow
regimes identified here, and beyond.

\section{Acknowledgments}

The \texttt{XSHELLS} code used for the numerical simulations is freely available at \url{https://bitbucket.org/nschaeff/xshells}.
We thank two anonymous reviewers for fruitful suggestions.
This work was funded by the French {\it Agence Nationale de la Recherche} under grant ANR-13-BS06-0010 (TuDy).
We acknowledge GENCI for awarding us access to resource Occigen (CINES) under grant x2015047382 and x2016047382.
Part of the computations were also performed on the Froggy platform of CIMENT (\texttt{https://ciment.ujf-grenoble.fr}), supported by the Rh\^ one-Alpes region (CPER07\_13 CIRA), OSUG@2020 LabEx (ANR10 LABX56) and Equip@Meso (ANR10 EQPX-29-01).
ISTerre is part of Labex OSUG@2020 (ANR10 LABX56).

\appendix
\section{Modal flows at $\Lambda_b = 0.190$}
\begin{figure}
  \includegraphics[width=\linewidth]{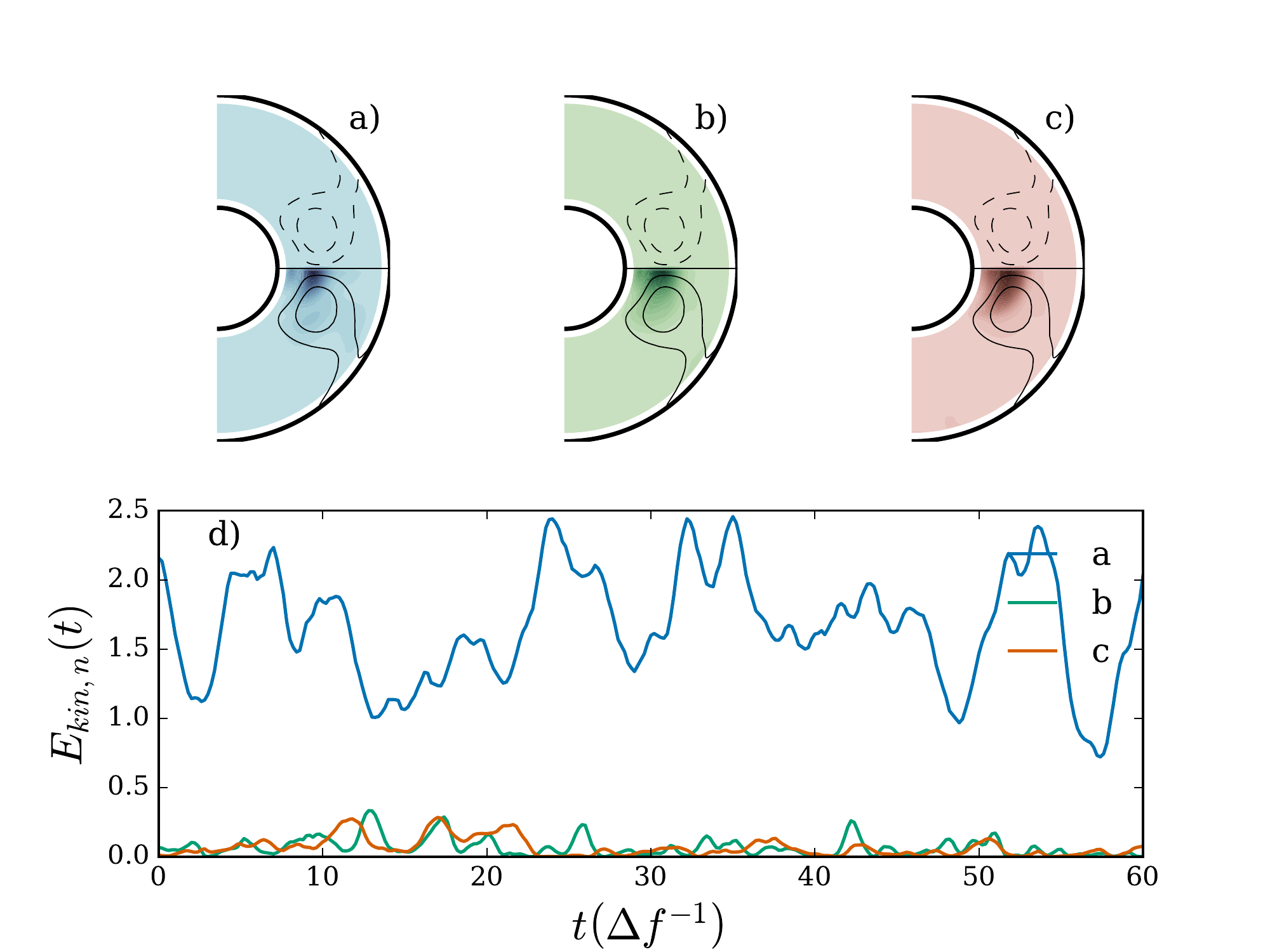}
  \caption{(a-c) Colormaps showing azimuthally averaged kinetic energy density of the
    first three singular modes of the $Ro = -1.0, \Lambda_b = 0.190$
    flow. The upper hemisphere shows the equatorially antisymmetric
    component of the mode; the lower hemisphere shows the equatorially
    symmetric component. (d) The energy in each mode (a-c) as a
    function of time, scaled against the mean energy in the kinetic
    fluctuations of the raw signal. Only the region $r \in [0.425,
      0.975]$ is included in the PCA. \label{fig:leftright-3}}
\end{figure}

\begin{figure}
  \includegraphics[width=\linewidth]{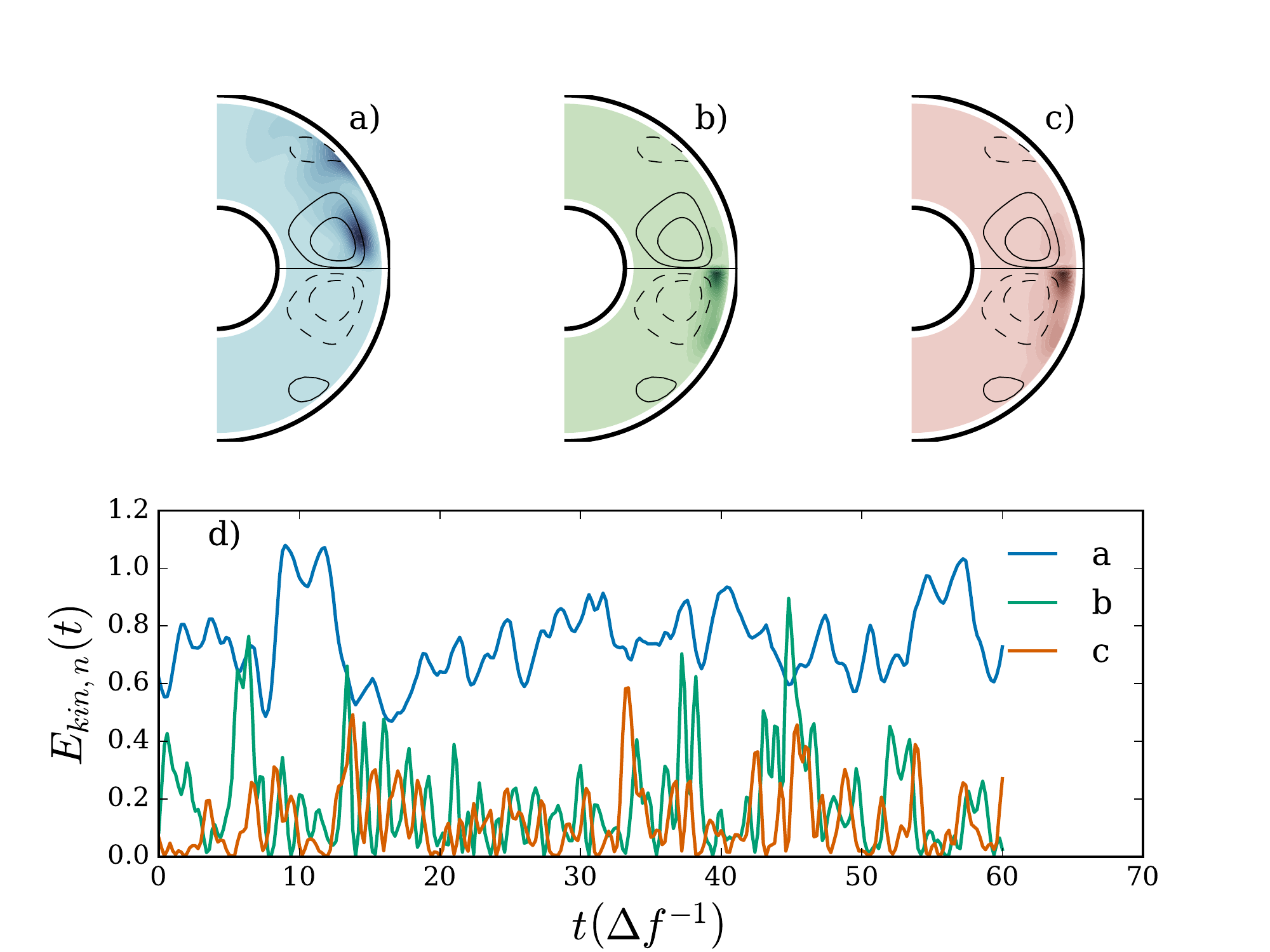}
  \caption{(a-c) Colormaps showing azimuthally averaged kinetic energy density of the
    first three singular modes of the $Ro = -1.2, \Lambda_b = 0.190$
    flow. The upper hemisphere shows the equatorially antisymmetric
    component of the mode; the lower hemisphere shows the equatorially
    symmetric component. (d) The energy in each mode (a-c) as a
    function of time, scaled against the mean energy in the kinetic
    fluctuations of the raw signal. Only the region $r \in [0.425,
      0.975]$ is included in the PCA. \label{fig:leftright-4}}
\end{figure}

As shown in Fig.~\ref{fig:dynregime}, simulations were carried out at
three values of $\Lambda_b$. Those discussed in the main text come
primarily from the case where $\Lambda_b = 0.095$, mostly to make
comparison easier between the modal and filamentary flows. The modal
structures and dominance of a single mode over the others is much more
clear in the case where $\Lambda_b = 0.190$. The decompositions of $Ro
= -1.0, \Lambda_b=0.190$ and $Ro = -1.2, \Lambda_b = 0.190$ are shown
in Figs.~\ref{fig:leftright-3}~\&~\ref{fig:leftright-4}
respectively. Note that the profiles of the various modes are very
similar to those of
Figs.~\ref{fig:leftright-1}~\&~\ref{fig:leftright-2}. The time series
show the primary difference between $\Lambda_b=0.190$ and
$\Lambda_b=0.095$, in Fig.~\ref{fig:leftright-4}, the antisymmetric
mode (a) is unambiguously the dominant mode (excepting a few bursts of
mode (b) where they reach equal energies).

\label{app:ModalFlows}

\section{Simulation Parameters}
\label{app:SimulationResolution}
The full simulations presented here were carried out using the
nondimensional parameters and resolutions shown in
Tab.~\ref{tab:resolution}. This table includes the magnetic Reynolds
$\left(Rm\equiv r_o^2\DOm/\eta = Ro / Em\right)$ for comparison with
other MHD simulations. Other works on MSC flow
\cite{Hollerbach.RSPA.2009} were characterized by the Hartmann number
$\left(Ha_{o,i} = r_{o,i} B_{o,i} / \sqrt{\mu \rho \nu \eta}\right)$ and
Reynolds number $\left(Re = r_o^2 \Delta\Omega / \nu\right)$,
specifically because these numbers do not require that the outer
sphere be rotating. Conveniently, the Ekman number is all that is
required to convert into our nondimensionalization the rotation
independent one $\left(Ha = \Lambda / \sqrt{Ek}\right.$, and $\left.Re
= \left|Ro\right| / {\rm Ek}\right)$. This gives $10^{5} < Re < 2 \times
10^6$. The Hartmann number varies between $6 < Ha_o < 18$ at the outer
sphere and $162 < Ha_i < 486$ at the inner sphere.  In the physical
experiment these values take the ranges $10^6 < Re < 2 \times 10^7$,
$Ha_o = 195$, and $Ha_i = 4430$.

\begin{table}
\begin{tabular}{llrrrrr}
\toprule
      &       &  $\Lambda_o$ &  $\Lambda_b$ &   $Em$ &  $Rm$ &  $\ell_{max}$ \\
$Ro$ & $f_o$ (Hz) &              &              &        &       &               \\
\midrule
-2.00 & 4.98  &        0.018 &        0.190 &  0.064 &  31.2 &           200 \\
      & 9.97  &        0.009 &        0.095 &  0.032 &  62.5 &           200 \\
      & 15.19 &        0.006 &        0.063 &  0.021 &  95.2 &           200 \\
-1.75 & 4.98  &        0.018 &        0.190 &  0.064 &  27.3 &           200 \\
      & 9.97  &        0.009 &        0.095 &  0.032 &  54.7 &           200 \\
      & 15.19 &        0.006 &        0.063 &  0.021 &  83.3 &           200 \\
-1.50 & 4.98  &        0.018 &        0.190 &  0.064 &  23.4 &           200 \\
      & 9.97  &        0.009 &        0.095 &  0.032 &  46.9 &           200 \\
      & 15.19 &        0.006 &        0.063 &  0.021 &  71.4 &           200 \\
-1.20 & 4.98  &        0.018 &        0.190 &  0.064 &  18.8 &           200 \\
      & 9.97  &        0.009 &        0.095 &  0.032 &  37.5 &           200 \\
      & 15.19 &        0.006 &        0.063 &  0.021 &  57.1 &           200 \\
-1.00 & 4.98  &        0.018 &        0.190 &  0.064 &  15.6 &           200 \\
      & 9.97  &        0.009 &        0.095 &  0.032 &  31.2 &           200 \\
-0.75 & 4.98  &        0.018 &        0.190 &  0.064 &  11.7 &           100 \\
-0.50 & 4.98  &        0.018 &        0.190 &  0.064 &   7.8 &           100 \\
      & 9.97  &        0.009 &        0.095 &  0.032 &  15.6 &           100 \\
 1.00 & 4.98  &        0.018 &        0.190 &  0.064 &  15.6 &           150 \\
      & 9.97  &        0.009 &        0.095 &  0.032 &  31.2 &           150 \\
 1.50 & 4.98  &        0.018 &        0.190 &  0.064 &  23.4 &           150 \\
      & 9.97  &        0.009 &        0.095 &  0.032 &  46.9 &           150 \\
\bottomrule
\end{tabular}
  \caption{Resolutions and nondimensional parameters of the
    simulations presented here. For all simulations the Ekman number
    $Ek = 10^{-6}$ , the time resolution is $10^{-4} \Delta f^{-1}$,
    and the radial grid has 384 points. This table includes the
    magnetic Reynolds $\left(Rm\equiv r_o^2\DOm/\eta\right)$ for
    comparison with other MHD simulations. The corresponding outer
    sphere rotation frequency $f_o$ in Hz of the DTS-machine is also
    indicated.
  \label{tab:resolution}}
\end{table}
\section{Principal Component Analysis of stable modes}
\label{appendix:PCA}
Because the flows discussed in Section~\ref{sec:pca} seem to primarily
be saturated instabilities, they're amenable to a principal component
analysis (PCA), wherein the flow is decomposed into individual spatial
modes, each with its own timeseries \cite{Abdi.CompStat.2010}. The
modes are chosen to maximize the cross correlation in time between
different measurements (here the kinetic and magnetic spectra at each
radial gridpoint ({\it i.e.} the $s_{\ell,m}(r_k), t_{\ell,m}(r_k),
S_{\ell,m}(r_k)$ and $T_{\ell,m}(r_k)$ of Eqn.~\ref{eqn:specdef}). The
analysis is carried in the frame of the inner sphere according to

\begin{equation}
  \mathbf{A} = \mathbf{L} \mathbf{\Sigma} \mathbf{R}^H.
\end{equation}

\noindent Here, $\mathbf{A}$ is a matrix representing spatio-temporal
data; $\mathbf{R}$ is a matrix of so-called right eigenvectors, each
representing a single spatial configuration that evolves together
($\mathbf{R}^H$ is the complex conjugate and transpose of the matrix
$\mathbf{R}$); $\mathbf{L}$ is a matrix of so-called left
eigenvectors, each representing the collective time evolution of its
corresponding right eigenvector; and $\mathbf{\Sigma}$ is a diagonal
matrix of singular values, each representing the relative strengths of
the modes.  The matrix $\mathbf{A}$ is computed by removing the means
of the magnetic and velocity fluctuations $\left(\widetilde{\bv}(r,
\theta, \phi, t_j)\; {\rm and} \;\widetilde{\uv}(r, \theta, \phi, t_j)
\;{\rm respectively}\right)$, and the singular value decomposition
(SVD) is carried out using \texttt{numpy}'s linear algebra module
\cite{SCIPY}.

The time series of Fig.~\ref{fig:surfpca}c and d are generated by
converting individual principle components $n$ back into a collection of
snapshots $\mathbf{u}_n\left(t_j\right)$ by 
\begin{equation}
  \mathbf{u}_n\left(t_j\right) = l_{j,n}\; \sigma_n\; \mathbf{r}_{n},
\end{equation}
\noindent where $\mathbf{r_n}$ is an individual column of
$\mathbf{R}$, $l_{j,n}$ is an individual entry in $\mathbf{L}$, and
$\sigma_n$ is the singular value of $n$.


\end{document}